\documentclass{sig-alternate-10pt}
\pdfoutput=1
\usepackage{fancyvrb}
\usepackage{fixltx2e}
\usepackage{bera}
\usepackage{graphicx}
\usepackage{epigraph}
\usepackage{subfig}
\usepackage{enumitem}
\usepackage[pdftex]{hyperref}
\title{Temporal and Spatial Classification\\
of Active IPv6 Addresses}

\begin{document}

\numberofauthors{2}

\author{
\alignauthor David Plonka \\
   \affaddr{Akamai Technologies} \\
   \email{plonka@akamai.com}
\alignauthor Arthur Berger\\
   \affaddr{Akamai Technologies \\ 
    Massachusetts Institute of Technology} \\
   \email{arthur@akamai.com}
}

\maketitle

\begin{abstract}
There is striking volume of World-Wide Web activity on IPv6 today.
In early 2015, one large Content Distribution Network handles 50
billion IPv6 requests per day from hundreds of millions of IPv6
client addresses; billions of unique client addresses are observed
per month.  Address counts, however, obscure the number of hosts
with IPv6 connectivity to the global Internet. There are numerous
address assignment and subnetting options in use; privacy addresses
and dynamic subnet pools significantly inflate the number of active
IPv6 addresses.  As the IPv6 address space is vast, it is infeasible to
comprehensively probe every possible unicast IPv6 address.  Thus, to
survey the characteristics of IPv6 addressing, we perform a year-long
passive measurement study, analyzing the IPv6 addresses gleaned from
activity logs for all clients accessing a global CDN.

The goal of our work is to develop flexible classification
and measurement methods for IPv6, motivated by the fact that its
addresses are not merely more numerous; they are different in kind.
We introduce the notion of classifying addresses and prefixes in two
ways: {\em (1)} {\em temporally}, according to their instances of
activity to discern which addresses can be considered stable;
{\em (2)} {\em spatially}, according to the density or sparsity of
aggregates in which active addresses reside.  We present measurement and
classification results numerically and visually that: provide details
on IPv6 address use and structure in global operation across
the past year; establish the efficacy of our classification methods;
and demonstrate that such classification can clarify dimensions of
the Internet that otherwise appear quite blurred by current IPv6
addressing practices.

\end{abstract}

\section{Introduction}		                        \label{sec:intro}
                                                        In 2015, we are in an era of production-quality, simultaneous operation
of the Internet protocol version 4 (IPv4) and version 6 (IPv6).
A number of observers have reported IPv6 traffic volume as doubling
in the past year, and globally over 6\% of clients having IPv6 connectivity~\cite{GoogleIPv6,HustonMarch2015,SOTI3Q14}.
In the fourth quarter of 2014, in some networks a significant proportion of World-Wide Web
(WWW) clients used IPv6 to access content that is available over both IPv6
and IPv4 via a global Content Distribution Network (CDN):
in the United States, this proportion was 70\% for Verizon Wireless, 30\% for
AT\&T, and 27\% for Comcast~\cite{SOTIIPv6}.

In this work, we study populations of active IPv6 addresses, {\em
i.e.,} those observed to be sources of traffic rather than merely
allocated or assigned.  Like most censuses, ours involves counting
members of groups or {\em classes.} IP addresses can be classified
with respect to various dimensions.  Historically, for IPv4, the initial address
``classes'' were determined {\em a priori} as classes A, B, C, {\em etc.}
Following the introduction of classless inter-domain routing
(CIDR), IPv4 addresses would more naturally be classified based on
flexible aggregates in routing tables, such as that of their Border
Gateway Protocol (BGP) prefix.  Addresses can also be classified based
on the set of reserved and special-use prefixes, {\em e.g.,} RFC1918
and multicast. However, operational needs have led to a broader notion
of address class, even if not referred to as ``class'' {\em per se,}
nor are classes mutually exclusive. Some example IPv4 address classifications
of recent interest are based on client reputation, geolocation,
assignment to a common network element ({\em e.g.,} router aliases),
anycast, and proxy.

Most of these classes pertain to both IPv4 and IPv6 addresses.
However, two dimensions are more significant with IPv6, and are the
focus of this paper.  The first we call ``temporal'' and is primarily
motivated by the popularity of host privacy extensions whereby the
vast majority of IPv6 addresses exist for short periods, {\em e.g.,}
24 hours or less, and in all likelihood will never be used again.
The second we call ``spatial'' and pertains to the vastly greater
number of possible areas (prefixes) and positions (addresses) in the IPv6 address space.
Whereas scanning the full IPv4
address space is now routine, this is not feasible for IPv6, and one
needs other techniques to discover ``where the action is.'' 
IPv6, also, allows greater freedom in the use of the subnet prefix.
We find a variety of practices employed by different network operators.
Our goal is to detect the different types of IPv6 addresses in active
use, with particular interest in $(a)$ discriminating {\em stable,}
persistent addresses from {\em ephemeral,} short-lived addresses and
$(b)$ discovering how addresses are arranged in the address space,
thereby forming {\em sparse} and {\em dense} regions.

There are numerous potential applications of temporal and spatial
address classification.  Examples include: selecting targets for
active measurements, {\em e.g.,} traceroutes, vulnerability scans,
and reachability surveys; informing data retention policy to prevent
resource exhaustion, {\em e.g.,} when encountering many ephemeral
addresses or prefixes; informing host reputation and access control,
{\em e.g.,} to mitigate network abuse; identifying homogeneous address
aggregates, {\em e.g.,} for IP geolocation; and detecting changes in
network operation or estimating Internet usage over time.

This paper makes the following contributions:\\ {\em (1)} We present
census results based on a large-scale, longitudinal, passive IPv6
measurement study of {\em active addresses} used by active
WWW clients in 133 countries and over four thousand autonomous systems.
{\em (2)} We introduce a {\em temporal} classification technique for
IPv6 addresses based on observation of address activity over time.
{\em (3)} We introduce a complementary {\em spatial} classification
technique for IPv6 addresses based on measurement of the sparsity or
density of the address prefixes in which they reside.  {\em (4)}
We evaluate the temporal and spatial classifiers by utilizing them
{\em in situ,} and show results of the classification of billions of
active IPv6 addresses.

In addition, we introduce the Multi-Resolution Aggregate
(MRA) plot, a visualization useful for examining populations of
addresses.  This plot is inspired by the work of Kohler {\em et
al.}~\cite{DBLP:conf/imc/KohlerLPS02}, and embellished for IPv6.
MRA plots show structural detail and allow address space exploration
without necessarily identifying specific addresses or blocks by
number.

Highlights of our measurement and classification results include,
as of early 2015:

\begin{itemize}[topsep=-1mm,wide,noitemsep]

\item When autonomous systems (ASNs) are ranked by their WWW client
address counts, the top 5 ASNs represent 85\% of active /64 prefixes
(``/64s'') and 59\% of all active addresses. Of these ASNs, 2 are
U.S.-based mobile carriers, {\em i.e.,} wireless Internet Service
Providers (ISP); the others are a European, an American, and a
Japanese ISP.

\item Although the vast majority of IPv6 clients use native transport,
6to4 tunneling is still common. If not segregated in measurement,
the ASNs hosting 6to4 relays would be amongst the top 5 ASNs.

\item 74\% of the 153 million of the /64s observed as active during two weeks
separated by 6 months are associated with just 1 ASN.

\item Despite the vast IPv6 unicast address space and generous
allocations to networks, many /64s are {\em reused}, {\em i.e.,}
assigned to different users over time, certainly within a week.

\item Of 1.81 million addresses observed as {\em stable} across 1 year,
over half a million are associated with two mobile carriers which,
in apparent contradiction, use {\em dynamic} values in network
identifiers.  Further investigation shows that many mobile devices
simultaneously use the same fixed interface identifier.  Combined with
dynamic /64 assignment, this can result in an IPv6 address being
reused by a different subscriber on a short timescale, {\em e.g.,} within days.

\item While privacy addressing is common and brings randomness and
sparsity to address values, there are many dense regions of IPv6
address space where addresses are well-ordered and tightly-packed.
49\% of active IPv6 ASNs have BGP prefixes containing such regions,
{\em e.g.,} /112 prefixes (64K address blocks) containing multiple
active WWW client addresses. These blocks are natural targets if
future, active scanning or probing is intended.

\end{itemize}

The remainder of this paper is organized as follows.  In
Section~\ref{sec:relwork}, we discuss related prior work. In
Section~\ref{sec:overview}, we give a brief introduction to IPv6
addresses.  In Section~\ref{sec:data}, we describe the data used in
our empirical study.  In Section~\ref{sec:method}, we present our
IPv6 address classification methods.  In Section~\ref{sec:results}, we
present results of our temporal and spatial classifications.
In Section~\ref{sec:discuss}, we discuss results and future work,
and subsequently conclude.

\section{Related Work}		                        \label{sec:relwork}
                                                        To our knowledge, our temporal classifier does not have a precedent
in the research literature.  The temporal characteristics of IPv4
addresses, however, became topical as scalability concerns arose with
the Internet's exponential growth in the 1990s.  Carpenter {\em et al.}
comment on this in RFC 2101~\cite{RFC2101}. With respect to IPv6,
Malone's work~\cite{DBLP:conf/pam/Malone08} is similar
to ours in that they also study active IPv6 addresses.
They develop a technique intended to classify short-lived privacy
addresses by examining only the address itself, but its accuracy is limited
by design, expected to identify approximately 73\% of all privacy addresses.
Since it is very challenging to detect randomness in short strings,
{\em e.g.,} 63 bits of an IPv6 address, we take
the complementary approach and identify those addresses that are stable
and, thus, almost certainly {\em not} privacy addresses.
In the end, Malone speculates that
``[accuracy] might be improved accounting for the times
addresses are observed and spatially/temporally adjacent addresses,''
which seems exactly the notion at which we arrive independently,
inspiring the strategies that we develop here.

Development of our spatial classifier is largely informed by the prior
work of two teams: Cho {\em et al.} and Kohler {\em et al.} Cho {\em et
al.}~\cite{DBLP:conf/qofis/ChoKK01} introduce {\em aguri,} a traffic
profiler that employs automatic aggregation based on addresses'
and prefixes' observed traffic volume.  As in their work, we find
their Patricia/radix tree-based aggregation useful in dealing with
resource constraints, however, we use it to discover {\em address
structure.}  We do this by aggregating to a threshold that is either
$(a)$ a percentage of total addresses or $(b)$ a prefix density,
rather than a percentage of total traffic volume.  This aggregation
method is useful because it generalizes to other metrics.

Kohler {\em et al.}~\cite{DBLP:conf/imc/KohlerLPS02} investigate the
structure of the IPv4 address space based on passive traffic analysis.
In a broad sense, our IPv6 investigation is similar and we employ
two of their metrics as-is: active aggregate counts and aggregate
population distributions.  IPv6 addresses, however, present different
challenges and opportunities to discern structure, so we develop new
IPv6-specific metrics.  Our work also differs in that we apply those
metrics to classify addresses rather than to evaluate mathematical
characterizations of the address space.

Dainotti {\em et al.}~\cite{DainottiCCR2014} investigate IPv4
address space usage by attempting to identify active and inactive /24
address-blocks using passive measurement. Our census of WWW client
addresses similarly employs passive means, but we count aggregates of
every possible prefix length.  Also, because we determine address
activity from the complete logs of all clients' successful WWW
transactions with a large CDN, we eschew complications introduced
by spoofed addresses.  While they propose that their method could
potentially apply to measuring IPv6 address space usage, they do not
discuss how it might treat persistent versus ephemeral addresses nor
if it could count addresses in ``small'' address-blocks, {\em e.g.,} smaller
than /64 prefixes.  Our method would likely complement theirs,
if applied to IPv6.

In 2012, Barnes {\em et al.}~\cite{BarnesISMA1202,KlaffyCCR2012}
evaluated methods to map the vast IPv6 address space by probes in order to
discover active addresses. Our work shares that goal but benefits
from increased IPv6 activity and content-accessibility
that make passive methods viable.  The stable addresses and dense
address regions that we identify are feasible targets for active scans
or probes, thus our method may repair or complement target
selection heuristics employed in their early survey. Still, like
Barnes {\em et al.,}
our work is guided by operator practice with respect to
IPv6 addressing.

There are a number of studies in the literature that measure and report
on the deployment and adoption of IPv6. Recent examples of such work
are those of Colitti {\em et al.}~\cite{DBLP:conf/pam/ColittiGKR10}
and Czyz {\em et al.}~\cite{Czyz2014} Our work differs in that we
measure IPv6 by counting active addresses and prefixes, rather
than by counting advertised prefixes or traffic
hits and bytes.  Each have different biases with respect to estimating usage.
Internet-wide surveys of active IPv6 addresses are scarce in the literature,
{\em e.g.,} Malone~\cite{DBLP:conf/pam/Malone08} circa 2008. However, Huston and
Michaelson~\cite{Huston2013,HustonMarch2015}, perform a
significant ongoing measurement study involving IPv6 addresses;
they observe activity by opportunistically running ``interactive''
advertisements that are crafted to elicit connection attempts
from WWW clients to their own measurement service via both IPv4
and IPv6. Our study is limited to IPv6, but seems to offer different
advantages.  They observe the IPv4/IPv6 address pairs associated
with WWW clients.  We observe significantly more activity from mobile
carriers, where the ads rarely run, and activity in a larger set of
ASNs~\cite{ggm2014,gih2015}.

\section{IPv6 Addresses}                       \label{sec:overview}

Here we present a brief introduction to IPv6 address assignment.
An IPv6 address consists of a leading network identifier, {\em a.k.a.}
subnet prefix, portion followed by an interface identifier (IID)
portion. The network identifier is used to route traffic destined
for this address to its Local Area Network (LAN) and the IID makes a
host interface's address unique on its local network segment.  While
superficially similar to the network and host identifier portions of
IPv4 addresses, the vast IPv6 address space allows much more freedom.

There are many IPv6 addressing schemes and network operators
are reminded to treat interface identifiers as semantically
opaque~\cite{RFC7136}. In this work, however, we utilize address
content, including IID, as a basis for classification and find
correspondences with a variety of standards-defined address types.
For instance, administrators have the option to use a /64 network
prefix and a rather large IID, {\em i.e.,} 64 bits~\cite{RFC4291}, or a
larger network prefix, {\em e.g.,} /127, and a smaller IID, {\em e.g.,}
only 1 bit~\cite{RFC3315,RFC6164}.  In the former case, with stateless
address auto-configuration (SLAAC), the host chooses
a 64-bit IID suffix for itself.  Consider the sample
addresses in Figure~\ref{fig:samplePresentation}.
In increasing order
of complexity, these addresses appear to be:
{\em (i)} an address with fixed IID value (\texttt{::103}),
{\em (ii)} an address with a structured value in the low 64 bits
        (perhaps a subnet distinguished by \texttt{::10}),
{\em (iii)} a SLAAC address with EUI-64 Ethernet-MAC-based IID, and
{\em (iv)} a SLAAC privacy address with a pseudorandom IID.

\begin{figure}[ht!]
\begin{quote}
\begin{scriptsize}
\begin{Verbatim}[commandchars=\\\{\},codes={\catcode`$=3\catcode`_=8}]
2001:db8:10:1:\textbf{:103}
2001:db8:167:1109:\textbf{:10:901}
2001:db8:0:1cdf:\textbf{21e:c2ff:fec0:11db}
2001:db8:4137:9e76:\textbf{3031:f3fd:bbdd:2c2a}
\end{Verbatim}
\end{scriptsize}
\end{quote}
\caption{Sample IPv6 addresses in presentation format with the low 64 bits shown bold.\label{fig:samplePresentation}}
\end{figure}

The first two addresses are similar to those created by traditional
addressing schemes used in IPv4 while the latter two use standard
IPv6-specific addresses schemes: EUI-64~\cite{RFC4862} and privacy
addresses~\cite{RFC4941}, respectively.~\footnote{Other IPv6 address
schemes by which interface identifiers are generated include:
Cryptographically Generated Addresses~\cite{RFC3972,RFC4982},
Hash-Based Addresses~\cite{RFC5535}, and stable privacy
addresses~\cite{RFC7217}.}
Since one might reasonably expect
these interface identifiers to be difficult to distinguish merely
by their content, we employ temporal analysis to discriminate these
from, at least, privacy addresses.

A number of transition mechanisms aid concurrent
operation of IPv6 with IPv4 and affect IPv6 addresses themselves. These
include: 6to4 relays~\cite{RFC3068} and Teredo~\cite{RFC4380}, which
employ global reserved prefixes; and ISATAP~\cite{RFC5214} which
embeds IPv4 addresses in the IPv6 IID.  Finally, there are additional
{\em ad hoc} schemes by which an IPv6 address contains an embedded
IPv4 address, {\em e.g.,} those used for some router and dual-stack
host interfaces. This is typically a convenience rather than
a requirement.

\section{Empirical Data}                            \label{sec:data}

Our study requires data sources containing
IPv6 addresses which are active, that is, addresses that exchange
globally-routed Internet traffic.

\subsection{WWW Client Addresses\label{subsec:logs}}

We primarily rely on aggregated logs of WWW server activity in this study.
These aggregated logs contain hit counts per client IP address.
We select only the client IP addresses from log entries that represent
successfully handled requests, thus avoiding spoofed sources.
The aggregation interval is 24 hours, for 55,000 of the CDN's IPv6-capable
servers, and is processed roughly by the end of the subsequent day.  
Note that the aggregation does not include the timestamp from the individual
log lines, used in separate processing for the CDN's customers.
Instead, we use the time epoch of the completion of processing
of the aggregated logs, which might be offset by as much as a day from when
the requests actually occurred.
Our stability analysis, described in Section~\ref{sec:method},
uses a heuristic to accommodate this timestamp slew.

In March 2015, the dataset contains IPv6 addresses in 6,872 BGP
prefixes originating from 4,420 ASNs (46\% of those advertising IPv6
prefixes). These figures are an increase from March 2014 when there
were 5,531 BGP prefixes originating from 3,842 ASNs (40\%).  Alas, we
certainly do not see traffic from all the world's WWW client addresses
at this observation point, and our stability analysis shows that some
specific long-lived active IPv6 addresses, {\em e.g.,} EUI-64, return
as WWW clients only infrequently.~\footnote{Some addresses that we
label as EUI-64 are false positives or have invalid or duplicate MAC
addresses, {\em e.g.,} MAC address {\tt 00:11:22:33:44:56} is the most
prevalent and just in one mobile carrier's network.
Otherwise, examination suggests these outliers are modest in number.}

\begin{table*}[ht!]
\scriptsize
\centering
\subfloat[Address characteristics per day]
{
\label{tab:dataDaySummary}
\begin{tabular}{|l||r|r|r|} \hline
{Characteristic}       & {Mar 17,}     & {Sep 17,}     & {Mar 17,}       \\
{}       & {2014}     & {2014}     & {2015}       \\
\hline\hline
Teredo addresses       & 1.98K (0.00\%)     & 3.28K (0.00\%)     & 20.1K (0.01\%)       \\ \hline
ISATAP addresses       & 90.2K (0.06\%)     & 101K (0.04\%)      & 133K (0.04\%)        \\ \hline
6to4 addresses         & 12.8M (7.97\%)     & 12.5M (5.90\%)     & 13.9M (4.19\%)       \\ \hline
{\bf Other addresses}  &{\bf 149M (92.0\%)}  &{\bf 199M (94.1\%)}  &{\bf 318M (95.8\%)} \\ \hline
\hline
{\bf Other /64 prefixes}&{\bf 61.4M}        & {\bf 82.9M}        & {\bf 121M}           \\ \hline
ave. addrs per /64        & 2.41               & 2.40               & 2.63                 \\ \hline
\hline
EUI-64 addr (!6to4)   & 3.13M (1.94\%)     & 3.66M (1.73\%)     & 4.49M (1.35\%)       \\ \hline
EUI-64 IIDs (MACs)     & 2.85M              & 3.23M              & 3.81M                \\ \hline

\end{tabular}
}
\subfloat[Address characteristics per week]
{
\label{tab:dataWeekSummary}
\begin{tabular}{|l||r|r|r|} \hline
{Characteristic}       & {Mar 17-23,}  & {Sep 17-23,}  & {Mar 17-23,}     \\
{}       & {2014}  & {2014}  & {2015}     \\
\hline\hline
Teredo addresses       & 15.1K (0.00\%)     & 24.5K (0.00\%)     & 131K (0.01\%)      \\ \hline
ISATAP addresses       & 210K (0.02\%)      & 238K (0.02\%)      & 346K (0.02\%)      \\ \hline
6to4 addresses         & 64.9M (7.22\%)     & 78.3M (6.34\%)     & 64.2M (3.43\%)     \\ \hline
{\bf Other addresses}  &{\bf 833M (92.8\%)} &{\bf 1.17B (94.9\%)}  &{\bf 1.80B (96.5\%)} \\ \hline
\hline
{\bf Other /64 prefixes}&{\bf 157M}           & {\bf 207M}         & {\bf 307M }         \\ \hline
ave. addrs per /64      & 5.32                & 5.64               & 5.88               \\ \hline
\hline
EUI-64 addr (!6to4)   & 8.88M (0.99\%)     & 13.1M (1.06\%)     & 16.2M (0.866\%)    \\ \hline
EUI-64 IIDs (MACs)     & 6.12M              & 8.16M              & 9.74M              \\ \hline

\end{tabular}
}
\caption{Active IPv6 WWW client address characteristics: March 2014 through March 2015.
\label{tab:dataSummary}
}
\end{table*}

Table~\ref{tab:dataSummary} summarizes the IPv6 WWW client address activity
observed across a year at 6 month intervals, March 2014 through March
2015; we report both daily and weekly counts.  
With daily counts, fewer ephemeral privacy addresses are observed,
while with weekly counts there is increased opportunity to observe
activity of WWW clients that visit the CDN less frequently than daily.

In Table~\ref{tab:dataSummary}, by March 2015, we see that the address
count increased to over 318 million observed daily and over 1.8 billion
observed in a week's time. Correspondingly, 121 million /64 prefixes
are observed daily and 307 million /64 prefixes in a week's time.

We are careful to separate client addresses involving some
IPv6 transition mechanisms from addresses involved
in ``native'' IPv6 end-to-end transport; this is because those
transition mechanisms' addresses would skew results. Specifically, we
cull addresses associated with the early IPv6 transition mechanisms,
{\em i.e.,} Teredo, ISATAP, and 6to4. Of these, only 6to4 still shows
significant use.
Since these 3 particular transition mechanisms' addresses are
easily classified, we focus our classifiers on the
``Other'' addresses, {\em i.e.,} those involving native,
end-to-end IPv6 transport.  Newer transition mechanisms such as
464-XLAT~\cite{RFC6877} and DS-Lite~\cite{RFC6333,RFC6674}, {\em e.g.,}
used by large mobile carriers, are included 
because they use IPv6 end-to-end, and thus represent native transport.
These ``Other'' addresses in Table~\ref{tab:dataSummary} account
for over 90\% of the active addresses observed.  Except for EUI-64
addresses, these can't easily be classified by
examination of address content based on standard formats.
These ``Other'' addresses are subjects for
the classifiers we introduce in Section~\ref{sec:method} and are
those for which we report results unless otherwise noted.

\subsection{Router Addresses\label{subsec:traceroute}}

In addition to periodic collection of active WWW client addresses, we
also collect a set of IPv6 addresses that were the source addresses
of ICMP ``Time Exceeded'' responses to our TTL-limited probes,
similar to those generated by the {\em traceroute} tool.
Based on collection in February 2015,
this dataset consists of 3.2 million addresses that appear to be
assigned to router interfaces.  Three types of probe targets
were used: {\em (1)} addresses of IPv6 recursive DNS servers, as
observed by our authoritative DNS servers, {\em (2)} addresses of
the CDN's servers in approximately 500 locations world-wide, and {\em
(3)} a selection of about 18 million WWW client addresses assembled
since 2013, including a subset (12 million) of those addresses
identified as stable in March and September, 2014 (Those reported in
Table~\ref{tab:dailyAddressStability} in Section~\ref{sec:results}.)
This dataset is used to identify additional dense prefixes as
reported in Table~\ref{tab:routerDensityCounts} with the expectation
that areas of the address space containing WWW clients differ from
those containing routers.

\section{Analysis Method} 		                \label{sec:method}
                                                        \subsection{\label{temporal}Temporal Classification}

Our temporal methods of IPv6 address classification are intended to
determine address lifetime, primarily to separate those client
addresses that are persistent or stable from those that are perhaps
not. We refer to this as {\em stability} analysis.  Let's first consider
a simple notion of stability.  If one periodically logs sets of active
addresses at some interval, {\em e.g.,} 6 months, it is easy to find
which sets have addresses in common. For instance, if address {\em
x} is observed as active in March 2015 as well as a year earlier,
in March 2014, it can be considered stable.  Our stability classes
are named according to the length of time across which stability
has been assessed. Thus we would say address {\em x} is ``1 year
stable,'' when sampled across the past 1 year, and is classified as
``1y-stable (-1y).'' If {\em x} is observed in March 2015 and also
6 months earlier, in September 2014, it would also be classified as
``6m-stable (-6m).'' This notion of stability generalizes to prefixes of
any length, not just full addresses; we similarly assess the stability
of /64 prefixes extracted from the full addresses.

Since we wish to perform stability analysis on an ongoing basis,
consider a slightly more complicated notion of stability.
Let's define more granular classes of stability, {\em e.g.,}
daily.  Definition: ``{\em n}d-stable'' is the class of addresses
for which there exist observations of activity on two different days
with an intervening time period of at least $n-1$ days.
For example, a given address seen on March 17 and again on March 18
(for which there are no intervening days) is said to be ``1d-stable.''
Likewise, an address seen on March 17 and on March 19 (for which there
is one intervening day) is said to be ``2d-stable.''  Note that 
since March 17 and March 19 have {\em at least} zero intervening days,
then an address seen on these two days is also ``1d-stable,'' besides being ``2d-stable;'' the classes are {\em not} mutually exclusive.
More generally, an address that is ``{\em n}d-stable'' is also ``{\em $(n-1)$}d-stable.''

Since a measurement study of client IP addresses that access a given
service will typically capture only a portion of the addresses' total
Internet activity, and since that service may be accessed infrequently,
even a long-lived client address, {\em e.g.,} using EUI-64, may
appear to be ephemeral.  Thus, we will simply label such addresses as
``not stable,'' meaning only that we do not
know that address to be stable.  Stability classification relies on,
and is limited by, the opportunity for observation of activity from
given vantage points.

In our daily stability analysis, we employ a sliding 15-day window
centered on the day of observation and spanning 7 days prior
through 7 days following.  In such context, a 3d-stable address
might be classified as ``3d-stable (-7d,+7d).'' For stability results
herein, ``(-7d,+7d)'' is implied unless otherwise noted.
Figure~\ref{fig:dataCountV6} in Section~\ref{sec:results} shows the
numbers of active addresses and /64s observed on each day as well
as the subset in common between those also observed on
the reference day (March 17 or March 23, 2015).

\subsection{\label{spatial}Spatial Classification}

Our spatial methods of IPv6 address classification and prefix
characterization are intended to both assess the proximity of addresses
and prefixes and to visualize the address blocks in which they are
contained.
We develop two related
metrics for use with IPv6: Multi-Resolution Aggregate (MRA) 
Count Ratios and Prefix Density, and a complementary visualization technique,
the MRA plot.  In the following, prefixes are characterized structurally, then
addresses therein are classified according to the densities of their
containing, non-overlapping sub-prefixes.

\subsubsection{\label{aggcount}Multi-Resolution Aggregate Count Ratios}

Our metric MRA Count Ratio is a generalization
of a metric introduced by Kohler {\em et al.} With 
128 bit addresses and IPv6 presentation format using hexadecimal
characters, network operators have great flexibility to use segments
of the address for internal purposes; {\em e.g.,} 16 bit and 4 bit
segments are commonly used for subnetting.  (See~\cite{BCOP_subnets}
and~\cite{IPv6addressPlan2013} for recommended operational guidelines.)
Here we present an informal, high-level understanding of MRA
ratios and an optional, formal definition.
The latter is unnecessary for a general introduction.

\begin{itemize}[topsep=-1mm,wide]

\item {\em Informally,} in the following MRA plots, the height
indicates how much that segment of the address is relevant to
grouping a set of addresses into areas of the address space.
Addresses aggregated further to the left (high order bits) are more
distant from each other; addresses aggregated to the right are close
to one another.  MRA ratios for a set of addresses, when plotted,
expose the density (or sparsity) of each segment of the addresses,
whether bits, characters, or colon-separated segments.

\item {\em Formally,}
Kohler {\em et al.} introduce the metric of active aggregate (prefix)
counts, and their ratio.  Given $N$ addresses, they
can be grouped into prefixes of various sizes.  For a given
prefix size, say $/p$, there is a (smallest) set of prefixes of size
$/p$ that contains (covers) all $N$ addresses.  At one
extreme, each IPv6 address is in its own $/128$ prefix, at the other
extreme the single $/0$ prefix contains all of the addresses.  Let the
``active aggregate count'' $n_p$ be the number of $/p$ prefixes that
covers the given set of addresses.  By definition $n_p = 1$ for
$p =0$ and $n_p = N$ for $p = 128$.  Often a more convenient metric
is the \emph{ratio} of active aggregate counts, $\gamma_p \equiv
n_{p+1} / n_p$.  The range of $\gamma_p$ is 1 to 2.  As an example,
suppose that a set of addresses is covered by 100 prefixes of size $/56$,
$n_{56} = 100$.  Now, consider one of these $/56$ prefixes and what can
happen when it is partitioned in two $/57$ prefixes. Either all of the
addresses in the $/56$ are in one of the $/57$ prefixes, or there is
at least one address in each of the two $/57$ prefixes.  If the former
pertains for {\em all} of the $/56$ prefixes then the ratio $n_{57} /
n_{56}$ would be 1, and if the latter pertains for all, the ratio would be 2.
Typically, the former pertains for some and
the latter for others, in which case the ratio is between 1 and 2.
Now, to examine 4-bit address segments, for instance, it is convenient
to compute ratios of active aggregate counts where the mask has
been incremented by values larger than 1 bit.  Note that 4 bits is
one hexadecimal character and 16 bits is a series of 4 hexadecimal
characters that, when aligned, are colon-delimited in IPv6 presentation
format; these are convenient segment sizes in IPv6 that are not
convenient in IPv4 due to presentation format being in base 10.
We consider the somewhat more general ``MRA count ratio''
$\gamma_{p}^{k} \equiv n_{p+k} / n_p$, where canonically
$p$ is a multiple of $k$, and $k$ is 1, 4, 8, or 16.  The range of
$\gamma_{p}^{k}$ is 1 to $2^k$.
Note: The definition of the ratios implies that,
for given resolution ($k$),
the product of the ratios is the total number of addresses in the set.

\end{itemize}

\begin{figure}[ht!]
\centering 
\subfloat[US university]
{
\label{fig:MRA_Uni}
\includegraphics[width=0.45\textwidth]{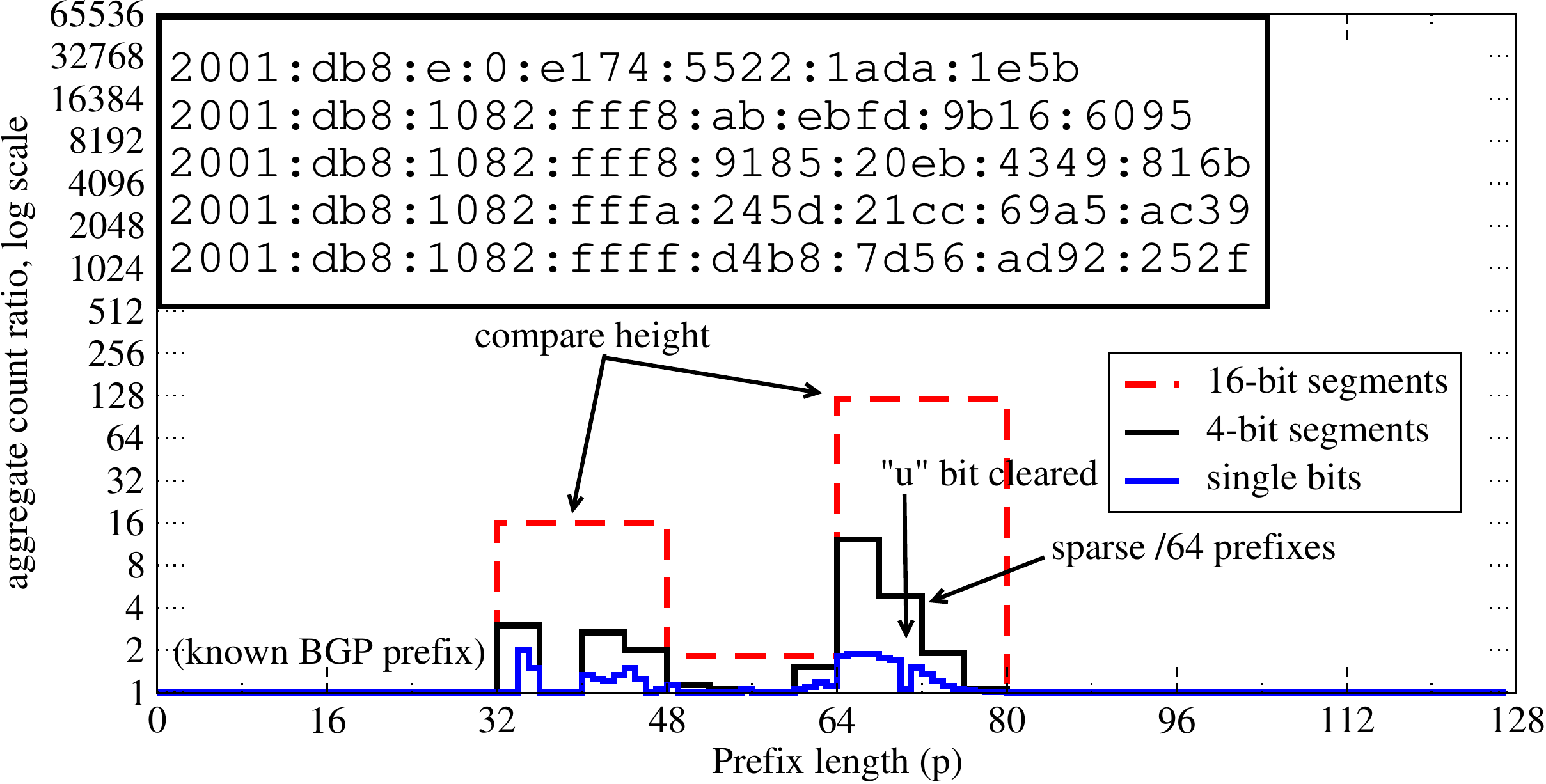}
}
\\
\subfloat[JP telco]
{
\label{fig:MRA_TelCo}
\includegraphics[width=0.45\textwidth]{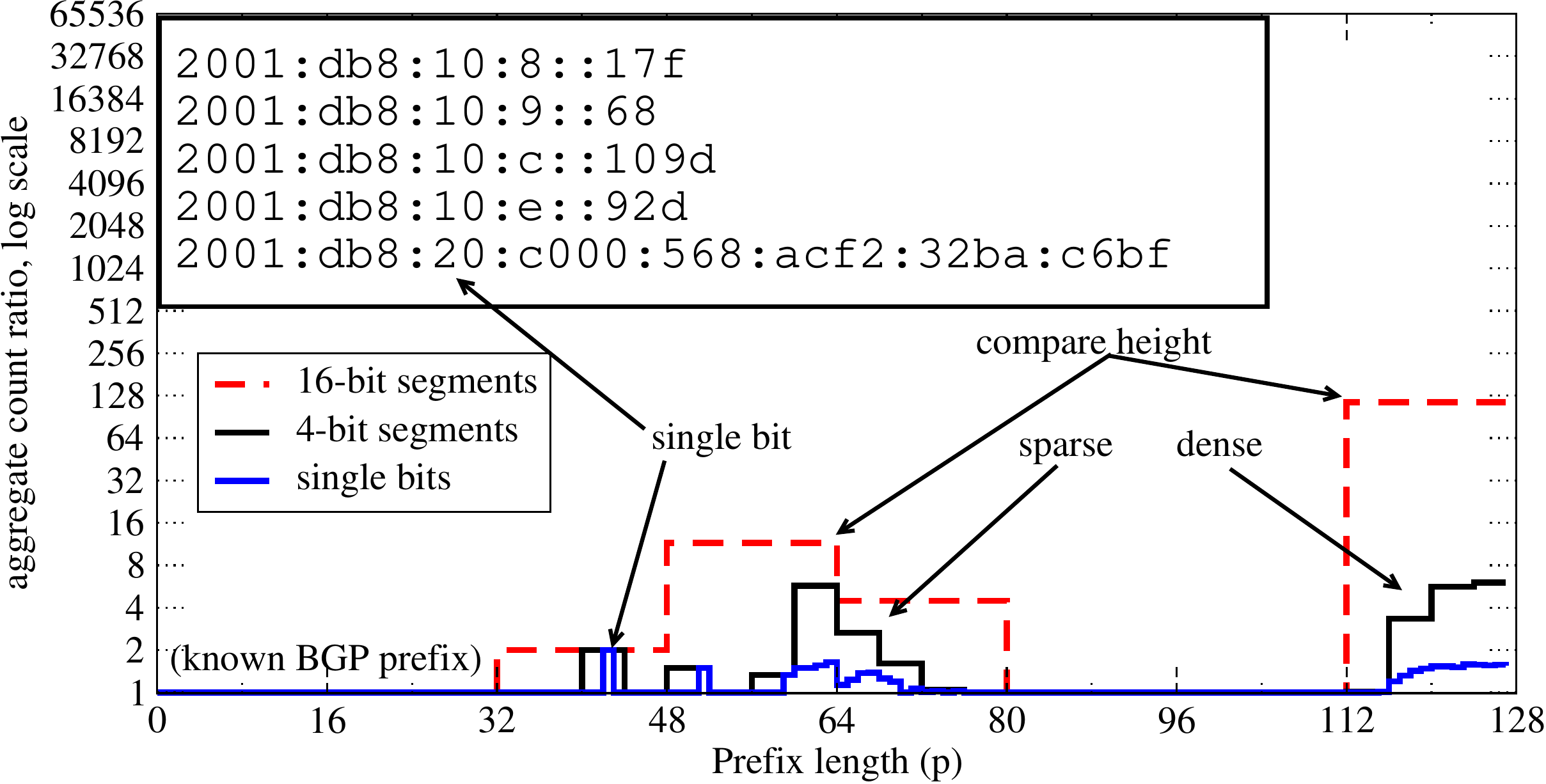}
}
\caption{MRA plots for active IPv6 WWW client addresses (a) 7.22K addrs and (b) 12.8K addrs.
\label{fig:SampleMRAPlots}
}
\end{figure}

Sample MRA plots are shown in Figure~\ref{fig:SampleMRAPlots},
annotated with sample addresses (inset) and arrows marking features
to aid the reader's interpretation.  In Figure~\ref{fig:MRA_Uni},
consider the portion of the plot for the ``single bits'' (blue) line
at $x >= 64$.  This portion of the plot initially approximates 2, then
slopes downward to the right, but with a drop to 1 at a particular
bit: this is the signature of a scenario where each /64 contains many
addresses and where, within each prefix, the majority of the addresses
have IIDs determined by the end host according to the pseudorandom
privacy extension as specified in RFC 4941~\cite{RFC4941}.

\begin{itemize}[topsep=-1mm,wide]
\item {\em Details on the signature for privacy extension:}
Consider one of the /64 prefixes.  Given that the
65th bit is chosen randomly, and that there are many addresses
in the /64 prefix, {\em e.g.,} $x$ addresses, then it is very likely 
(probability = $1 - (1/2)^{x-1}$) 
that at least one
of the addresses will have a 0 for the 65th bit and, likewise, that at least
one of the addresses will have a 1. Thus the ratio $n_{65}/n_{64}$
for this prefix is very likely to be 2.  If this pertains for
all /64 prefixes, the ratio for the whole set of addresses
will also be 2.  In turn, given that each /65 prefix also has a
large number of addresses, the above logic repeats, and thus we expect
the ratio to remain close to 2.  However, as we continue to split
prefixes in half, each prefix has a decreasing number of addresses and
an increasing chance that those addresses will all have the same next bit.
Once they are all the same, the ratio for such prefixes will be
1 and the overall ratio will begin to decline from 2.  Moreover,
even if the original set contained a billion addresses,
it would still be very sparse in the space of $2^{64}$ possible IIDs.
As we continue to consider ever smaller prefixes, eventually,
each will contain just one pseudorandom-IID address,
and the overall ratio will flat line at 1.  In the presented
plot, this occurs at about the 80th bit.  Finally, as a defining feature
of the present scenario, note that the ratio drops to almost 1 at
the 71st bit, shown at 70 on the horizontal axis.  This is consistent
with end hosts that determine the IID according to RFC 4941,
which specifies that the ``u'' bit be set to 0, meaning that an IID
is not necessarily universally unique, as opposed to a MAC address.
\end{itemize}

\vspace{3mm}

Now consider Figure~\ref{fig:MRA_TelCo} in contrast to
Figure~\ref{fig:MRA_Uni}.  These two organizations appear to
have significantly different address assignment policies.  In
Figure~\ref{fig:MRA_TelCo}, we see a prominence between bits 112 and
128. This indicates that there are many active addresses that differ in
only those least-significant bits, {\em i.e.,} addresses are clustered within
smaller prefixes, and thus such prefixes are more dense address blocks.

If one were interested in searching for additional active IPv6
addresses, these denser address blocks would be natural targets.  A /112
prefix covers $2^{16}$ addresses, the same as a /16 in IPv4, and is
easily scanned, whereas scanning across a /64 is not practical.

Now consider Figure~\ref{fig:MRA_Uni} with respect to the
plotted ratios for 4-bit segments, {\em a.k.a. nybbles}, in the
plot (black line).  Here we consider changes on a per-nybble,
hexadecimal character basis.  This provides a more aggregated view,
summarizing details of changes on a per-bit basis.  Our first-order
interest is network operator practice with respect to subnetting.
In particular, we assume this network subnets their /32 BGP prefix,
so we consider segments of the address down to the /64, {\em i.e.,}
across the canonical network identifier.  The jump up for the plot
at 32 indicates that addresses have differing (character) values
at that nybble, but not, in turn, at the subsequent nybble at 36.
The subsequent two nybbles could also be used to discriminate many
addresses, and then much less so for the subsequent three nybbles.
In contrast to Figure~\ref{fig:MRA_Uni}, note that the addresses in
Figure~\ref{fig:MRA_TelCo} have many different (character) values
in the last nybble within network portion of the address at 60.
In Section~\ref{sec:results}, {\em e.g.,} Figure~\ref{fig:MRAPlots},
we examine aggregation ratio across the advertised BGP prefixes,
{\em i.e.,} operator-defined address blocks in the unicast portion
of the IPv6 address space.

While this introduction to Multi-Resolution Aggregation ratio focussed
on visual recognition of IPv6 features in MRA plots, the underlying
$x,y$ values offer a convenient basis to classify prefixes, and the
addresses therein. While defining MRA-based address
classes is left for future work, we begin by 
developing spatial classification by identifying dense prefixes.

\subsubsection{Prefix Density}

Kohler {\em et al.}~\cite{DBLP:conf/imc/KohlerLPS02} introduces the
metric of the number of active addresses in a prefix, and examine
the distribution of this number across prefixes of a given size.
They consider IPv4 and are interested in the variability of
population densities for prefixes of a given size, {\em e.g.,} /8 or /16,
and how well their models match with measurements.  They comment, ``aggregate
population distributions are the most effective test we have found
to differentiate address structures.''

\begin{figure}[h!]
\centering
\includegraphics[width=0.45\textwidth]{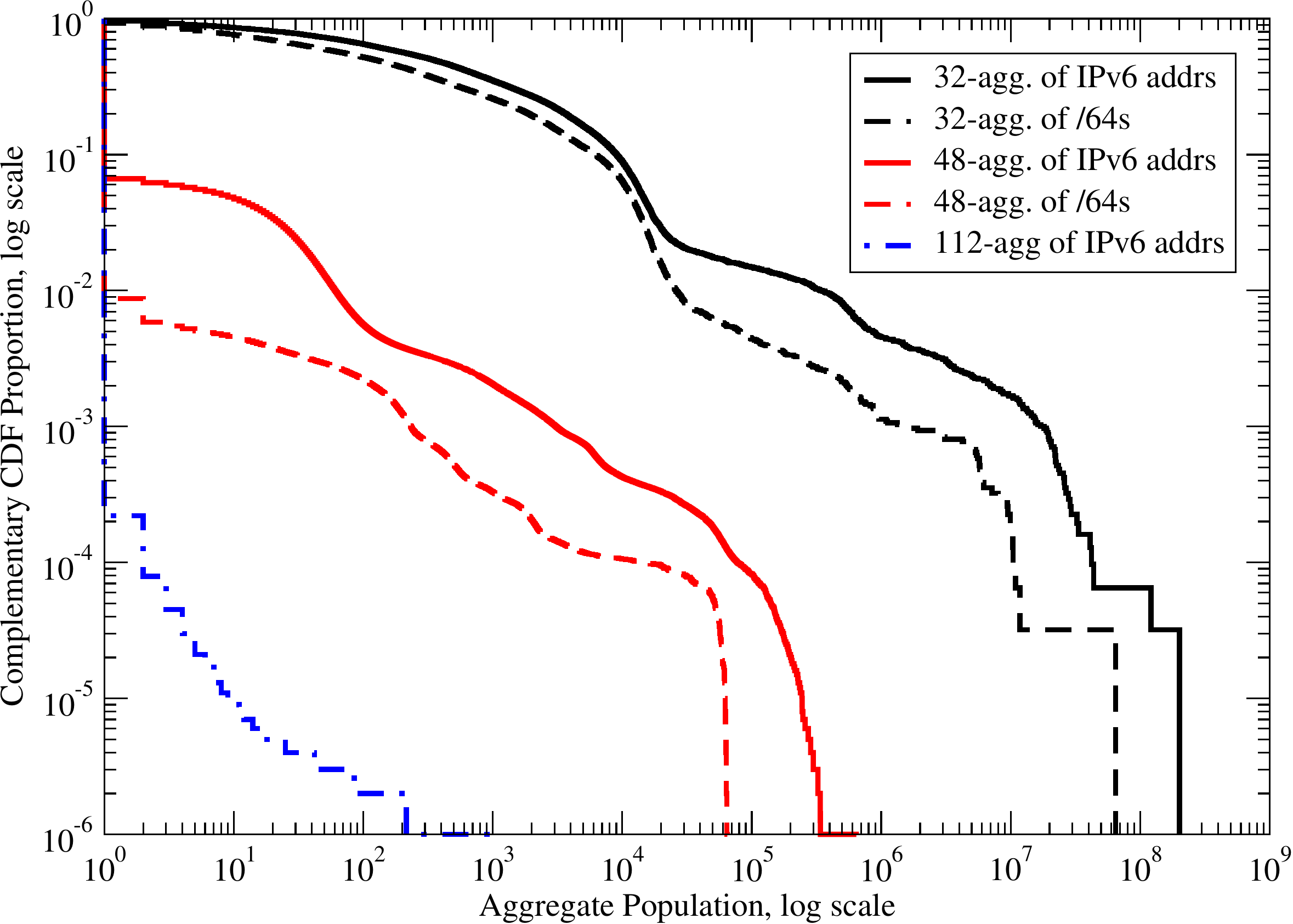}
\caption{Aggregate population distributions for 1.87B
IPv6 addrs, 358M /64s, March 17-23, 2015.
\label{fig:aggPopDist}
}
\end{figure}

We plot the aggregate population complementary cumulative distribution
function for
all IPv6 addresses and /64 prefixes active during a 7-day period
in~Figure~\ref{fig:aggPopDist}.  For the curve showing the 112-aggregate
of addresses (the lowest curve), only $10^{-5}$ of the /112 prefixes
contained 10 or more observed addresses; for the 48-aggregate of
addresses, fewer than one in ten of the /48 prefixes contained 10 or
more observed addresses, hence, a few prefixes must contain most of
the addresses. Approximately $10^{-4}$ of the 48-aggregate of addresses contain
$10^5$ or more addresses, which clearly illustrates the sparsity of
the IPv6 address space and the concentration of observed addresses
in a small subset of prefixes.

Kohler's aggregate population considers the observed {\em count}
of addresses in a prefix.  A related measure is obtained by dividing
that observed count by the number of addresses spanned by the prefix,
yielding the {\em percentage} of the addresses of
the prefix that were observed.

Cho {\em et al.}~\cite{DBLP:conf/qofis/ChoKK01} use a percentage as a
criterion in an aggregation-based traffic profiler, {\em however,} their
percentage is obtained by dividing an observed count by a total observed
count across all prefixes. In their implementation, prefixes are nodes
in an {\em aguri tree,} with observed addresses added as leaf
nodes. Aggregation is kind of ``pruning,'' and is performed by
aggregating a node's count to its parent (and removing that node),
unless that node's count meets or exceeds a target minimum percentage.

Consider an IP network prefix such as {\tt 2000::/3} or {\tt
2001:db8::/32}.  A prefix might contain ``addresses of interest'' by
arbitrary criterion, {\em e.g.,} addresses for which activity was observed.
A simple notion of a prefix's {\em density}, then, is the
fraction of its addresses that are active. Both the prefix and
its addresses can be said to have a density of {\em d}. where
{\em d} is a fraction with a value greater than 0 and less than
or equal to 1.

If we restrict desired minimum densities to the fraction
$n/2^{p'}$, where $p'$ is a number of bits in the range $0$ through $128$,
there is a simpler solution that does
not require base-10 math with large numbers, {\em i.e.,} greater than
64 bits. We use this restriction, and choose densities based on two parameters:
$n$ and $p$, where $p = 128 - p'$.

Now, let's define our spatial address classes based on prefix density.
Definition: ``$n$@$/p$-dense'' is the class of prefixes of length $p$
that contain at least $n$ addresses for which there exist observations
of activity. It is also the class of those addresses contained therein.
For example, let's say the IPv6 addresses {\tt 2001:db8::1} and {\tt
2001:db8::4} are both active, but no others. If the desire is to identify
/112 prefixes that are dense, then {\tt 2001:db8::/112} is the sole
$2$@/$112$-dense prefix. There is also one $2$@/$125$-dense prefix, but no
$2$@/$126$-dense prefixes.

\subsubsection{\label{computingDense}Computing Dense Prefixes}

We would like to identify the dense prefixes based on observed
active addresses.  We start by choosing a desired minimum density,
and then compute the set of dense prefixes, if any.

Given a set of IP addresses and a desired minimum density, we compute
a corresponding set of prefixes that {\em (a)} contain a subset of
those addresses, {\em (b)} have the desired density,
and {\em (c)} have prefix length
up to 127.  The dense subset are the least-specific, non-overlapping
prefixes that are dense, {\em i.e.,} contain the requisite fraction
of addresses.

One way to implement this ``densification'' is by using an aguri
tree,~\cite{DBLP:conf/qofis/ChoKK01} (a base-2 radix tree, {\em a.k.a.,}
Patricia trie) augmented with a new ``densify'' operation that works as follows:

\begin{enumerate}[topsep=-1mm,wide,noitemsep]

\item[{\em (1)}] Populate the tree by adding each of the addresses
with a count of 1.  If dense prefixes of just that one length
are desired, add each address with a ``/$p$'' and skip to step {\em
3}.~\footnote{When prefixes of just one length are desired, the aguri
tree is unnecessary; it is just accumulating counts and sorting output.
An alternative is to print addresses in a fixed-width 32-character
hex format, one per line, and use:
{\tt sort [-m] |cut -c1-\$(($p$/4)) |uniq -c}~\cite{GNUtextutils}}

\item[{\em (2)}] Perform a post-order traversal of the tree; when visiting
a node that has children and the sum of counts for the current node
and its children would make the current node's prefix of the desired
density, aggregate the node's children into the current node, by
accumulating the count and removing the children.

\item[{\em (3)}] Now the least-specific dense prefixes, of at least
the desired length $p$ (if specified), are nodes in the tree.  However,
addresses in sparse regions remain unaggregated, so they are present as
well, {\em e.g.,} /128s.  To report only the dense prefixes, perform
an in-order traversal, skipping those with a count that is less than
$n$, {\em e.g.,} 2, and print others as they are dense prefixes.

\end{enumerate}

\section{Results}                                       \label{sec:results} 

\subsection{Temporal Classification}

Table~\ref{tab:stabilityResults} summarizes the
temporal classifications for the ``Other'' addresses in
Table~\ref{tab:dataSummary} of Section~\ref{sec:data}.

\begin{figure*}[ht!]
\centering
\subfloat[IPv6 address stability]
{
\label{fig:dataCountV6addrs}
\includegraphics[width=0.5\textwidth]{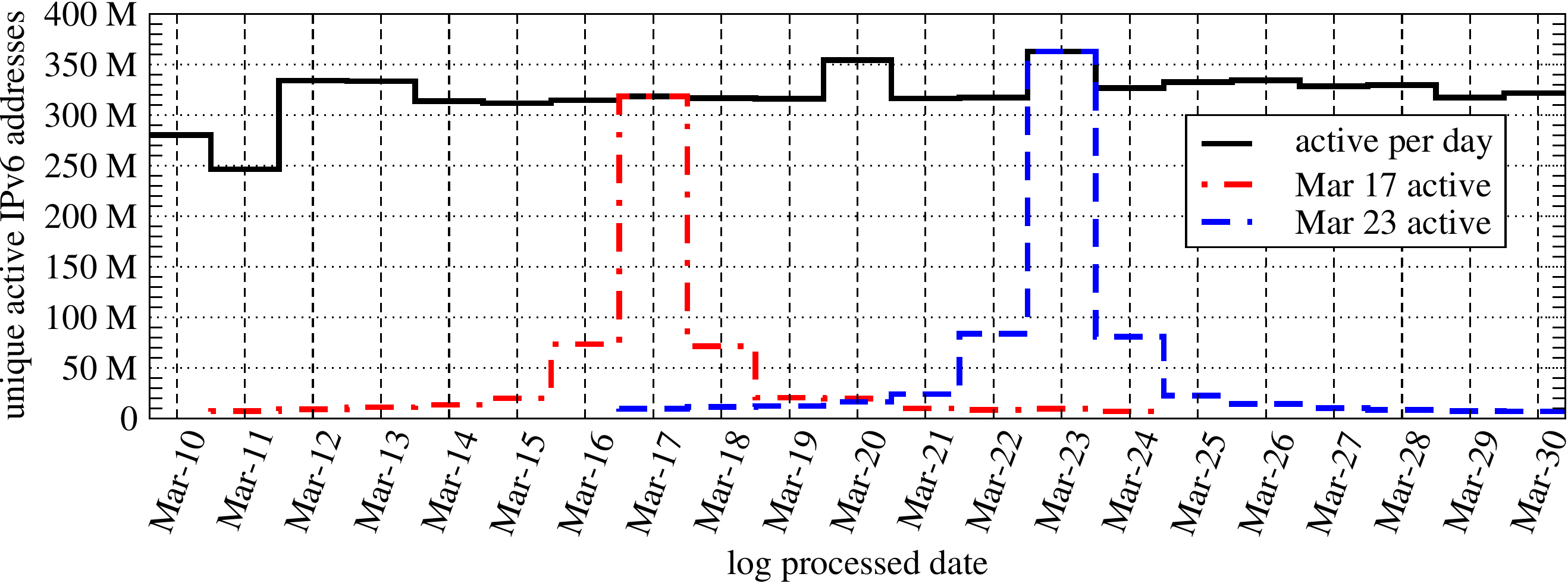}
}
\subfloat[/64 prefix stability]
{
\label{fig:dataCountV6_64}
\includegraphics[width=0.5\textwidth]{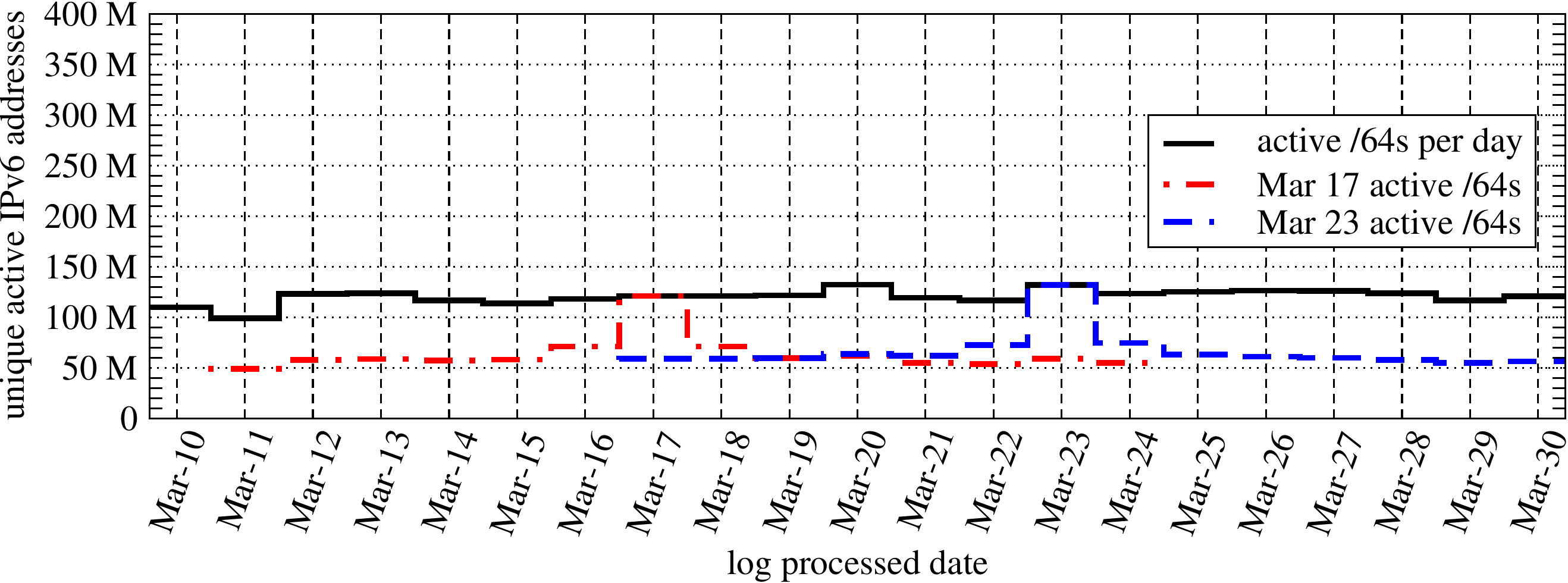}
}
\caption{Stability study of active IPv6 WWW client addresses and prefixes observed per day, March 2015.~\label{fig:dataCountV6}}
\end{figure*}

Based on our temporal classification method as described in
Section~\ref{temporal}, Figure~\ref{fig:dataCountV6} shows 
stability of active addresses and /64 prefixes observed on March
17 and 23, 2015, by 15-day sliding window.  Consider the values for
``March 17 active'' (red) in Figure~\ref{fig:dataCountV6addrs}.
Here we see that about 320 million WWW client IPv6 addresses were
observed on March 17. Of those addresses, about 75 million were
also seen the previous day, about 20 million the day before that,
about 10 million the day before that, and so on, in stepwise fashion.
The same is true, approximately symmetrically, for the days
following March 17.  Ultimately, this assessment yields 30.1
million 3d-stable addresses (9.44\%), as listed in  the ``Mar 17,
2015'' column of Table~\ref{tab:dailyAddressStability}.

Now consider the corresponding stability of /64 prefixes shown
in Figure~\ref{fig:dataCountV6_64}.  A larger proportion of
/64 prefixes are stable than that of full addresses: 109 million
3d-stable /64s (89.8\%), as listed in the ``Mar 17, 2015'' column of
Table~\ref{tab:daily64Stability}.  (The upper limit on the number
stable addresses is the number of stable /64s, or stable prefixes
of any length.)

See Tables~\ref{tab:weeklyAddressStability}
and~\ref{tab:weekly64Stability} for the stability results of addresses
and /64s, respectively, over a week's time.  For each of the seven
days, the 3d-stable addresses are determined, and the table reports
the count of the unique 3d-stable addresses seen over those days.
Likewise for the ``not 3d-stable.''

On examining the values highlighted (bold) in
Table~\ref{tab:stabilityResults}, we make two notes: {\em (a)} in a
relative sense, there are not many very long-lived WWW client IPv6
addresses, only 1.81 million (0.1\%) observed over the course of
a year; and {\em (b)}, there are many long-lived /64 prefixes for
active WWW clients: 153 million 6m-stable /64s, and even 116 million
1y-stable /64s.
Consider Figure~\ref{fig:bgpCounts}, where we plot the CCDF of
various counts by ASN.  We see that a single
ASN accounts for over 100 million /64s (dashed black) as observed across 6 months,
indicating that most long-lived /64s (dashed blue) are in only a few networks. We
explore this further in Section~\ref{MRA}.

\subsubsection{Discussion of Temporal Results}

One motivation for identifying 3d-stable addresses is the proposal
that they would be good targets for subsequent active probing to discover network infrastructure.
We tested this hypothesis by using a randomly selected subset of 3d-stable IPv6
addresses as targets for TTL-limited probes.  We discovered 129\% (1.8
million) more active IPv6 router addresses than using a simpler, long-standing
target-selection strategy that works well with IPv4. (The
IPv4 strategy is based only on selecting target addresses of recursive
name servers that query the CDN's authoritative servers and randomly
selected addresses of active WWW clients.)

As for the ``not 3d-stable'' addresses, we expect that the vast majority are hosts using a privacy-extension
IID, as the default timeout is 24 hours.~\cite{RFC4941}
However, other types of addresses are present as well.  
Note that, although the IID of EUI-64 addresses is static,
the subnet prefix can vary, as when the device is moved between networks, or when a given operator implements
a policy of assigning another subnet prefix each time the device connects to the network.  (See Section
\ref{MRA} for further discussion.)
We investigated EUI-64 addresses in the Sept. 17-23, 2014 dataset that were classified as ``not 3d-stable.''
In 62\% of them, the IID appeared in more than one address.
Also, for 14\% of them, the IID also appeared in an address that was classified as 3d-stable.

While the temporal class ``3d-stable (-7d,+7d)'' is useful when
applied to target selection for active measurements, more research is
warranted in order to determine what specific temporal classes may be most useful,
{\em e.g.,} varying the number of days or the sliding window size,
and in combination with addressing practices.

One wonders whether or not any of our counts of active or stable
/64s could be an approximate lower-bound on the actual number of
subscribers or instances of IPv6-capable Internet connections
in the world today.  Consider the
highlighted (bold) counts in Table~\ref{tab:weekly64Stability}; 
is the 116 million 1y-stable /64s
observed March, 2015, a reasonable lower-bound?  We discuss this in
the forthcoming Section~\ref{sec:discuss}, but first, we 
address spatial classification results in order to better understand how
network identifiers such as /64 prefixes are assigned by ISPs.

\begin{table*}[ht!]
\scriptsize
\centering
\subfloat[Stability of IPv6 addresses per day]
{
\label{tab:dailyAddressStability}
\begin{tabular}{|l||r|r|r|} \hline
{addr class} & {Mar 17, 2014}    & {Sep 17, 2014} & {Mar 17, 2015}\\
\hline\hline
3d-stable        & 13.7M (9.22\%)& 13.6M (6.84\%) & 30.1M (9.44\%)\\ \hline
not 3d-stable    & 134M (90.8\%) & 185M (93.2\%)  & 288M (90.6\%)\\ \hline
\hline
6m-stable (-6m)  & {}                & 588K (.296\%)  & 1.08M (.340\%)\\ \hline
1y-stable (-1y)  & {}                & {}             & 328K (.103\%)\\ \hline
\end{tabular}
}
\subfloat[Stability of /64 prefixes per day]
{
\label{tab:daily64Stability}
\begin{tabular}{|l||r|r|r|} \hline
{/64 class}      & {Mar 17, 2014} & {Sep 17, 2014} & {Mar 17, 2015}\\
\hline\hline
3d-stable        & 55.8M (91.0\%) & 74.6M (89.9\%) & {\bf 109M (89.8\%)}\\ \hline
not 3d-stable    & 5.53M (9.01\%) & 8.33M (10.1\%) & 12.3M (10.2\%)\\ \hline
\hline
6m-stable (-6m)  & {}             & 23.4M (28.2\%) & 32.4M (26.7\%)\\ \hline
1y-stable (-1y)  & {}             & {}             & 21.8M (18.0\%)\\ \hline
\end{tabular}
}
\\
\subfloat[Stability of IPv6 addresses per week]
{
\label{tab:weeklyAddressStability}
\begin{tabular}{|l||r|r|r|} \hline
{addr class}    & {Mar 17-23, 2014} & {Sep 17-23, 2014}  & {Mar 17-23, 2015} \\
\hline\hline
3d-stable          & 37.0M (4.44\%) & 34.0M (2.91\%)     & 69.0M (3.82\%)\\ \hline
not 3d-stable      & 796M (95.6\%)  & 1.13B (97.1\%)     & 1.74B (96.2\%)\\ \hline
\hline
6m-stable (-6m)    & {}                & 3.25M (.280\%)     & 3.66M (.202\%)\\ \hline
1y-stable (-1y)    & {}                & {}                 & {\bf 1.81M (.100\%)}\\ \hline
\end{tabular}
}
\subfloat[Stability of /64 prefixes per week]
{
\label{tab:weekly64Stability}
\begin{tabular}{|l||r|r|r|} \hline
{/64 class}        & {Mar 17-23, 2014} & {Sep 17-23, 2014}  & {Mar 17-23, 2015} \\
\hline\hline
3d-stable          & 131M (83.7\%)     & 169M (81.8\%)      & {\bf 246M (80.3\%)}\\ \hline
not 3d-stable      & 25.5M (16.3\%)    & 37.7M (18.2\%)     & 60.6M (19.7\%)\\ \hline
\hline
6m-stable (-6m)    & {}                & 120M (58.1\%)      & {\bf 153M (49.9\%)}\\ \hline
1y-stable (-1y)    & {}                & {}                 & {\bf 116M (37.8\%)}\\ \hline
\end{tabular}
}
\caption{Stability of active IPv6 WWW client address and prefix counts, not 6to4 or Teredo, March 2015.
\label{tab:stabilityResults}
}
\end{table*}

\subsection{Spatial Classification}

\subsubsection{\label{MRA}Multi-Resolution Aggregate Count Ratio}

\begin{figure*}[ht!]
\centering
\subfloat[Distribution of active addrs and /64 counts, 4.42K ASNs]
{
\label{fig:bgpCounts}
\includegraphics[width=0.50\textwidth]{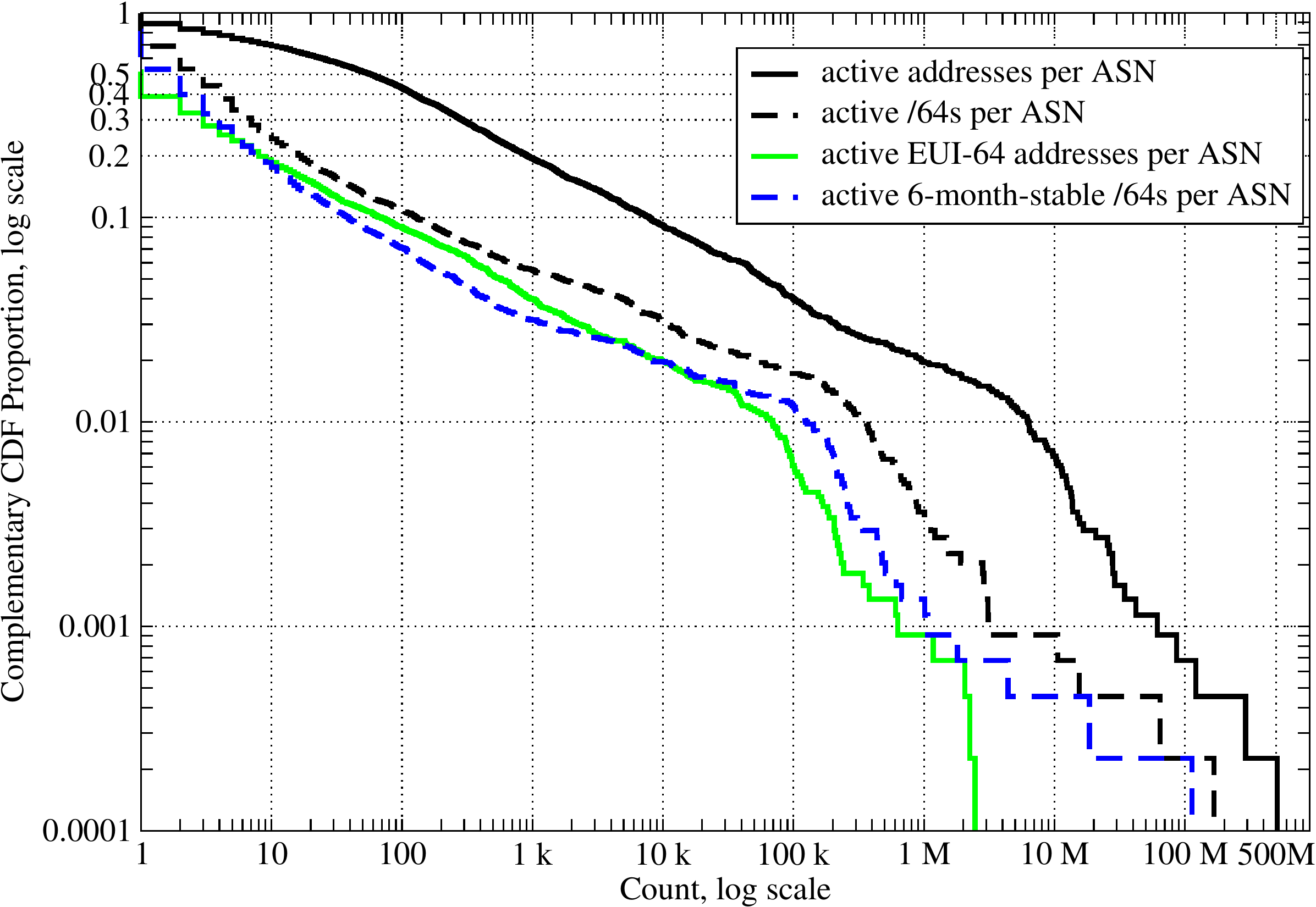}
}
\subfloat[16-bit segment agg. distributions, 6.87K BGP prefixes]
{
\label{fig:MRAdist}
\includegraphics[width=0.50\textwidth]{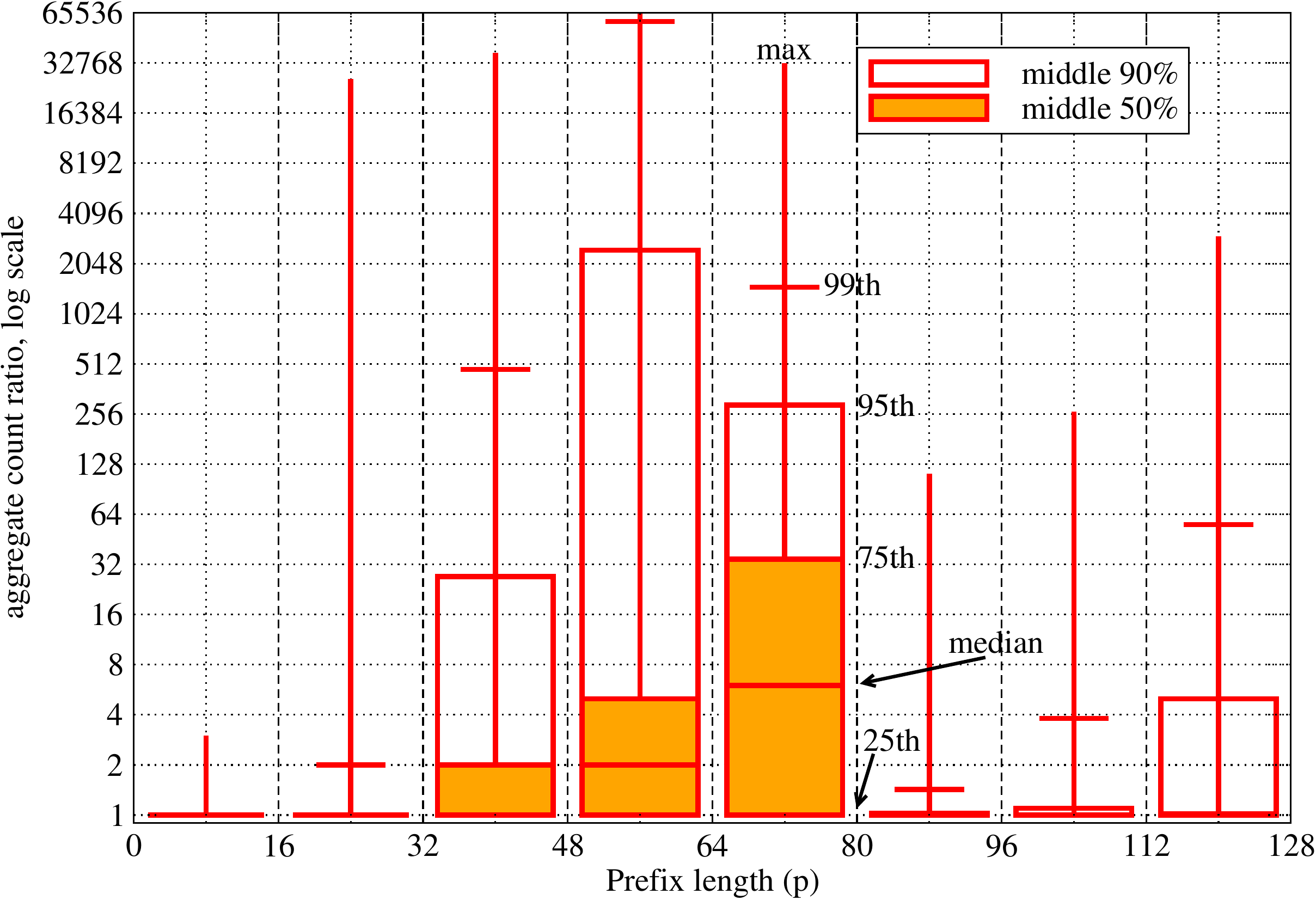}
}
\\
\subfloat[All: 1.81B active IPv6 client addrs, not 6to4 or Teredo]
{
\label{fig:allMRAPlot}
\includegraphics[width=0.50\textwidth]{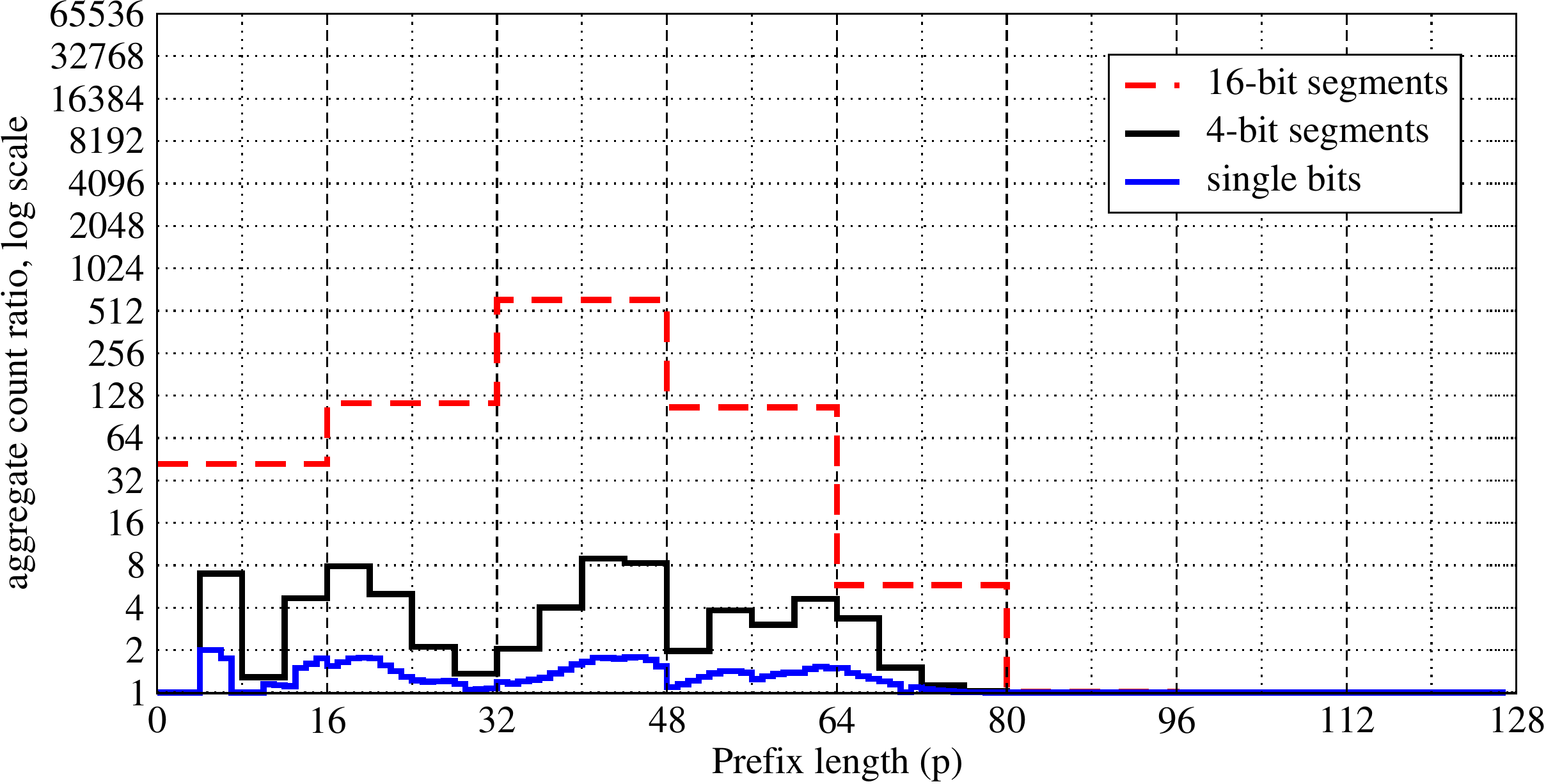}
}
\subfloat[6to4: 64.2M active IPv6 (49.3M IPv4) client addresses]
{
\label{fig:6to4MRAPlot}
\includegraphics[width=0.50\textwidth]{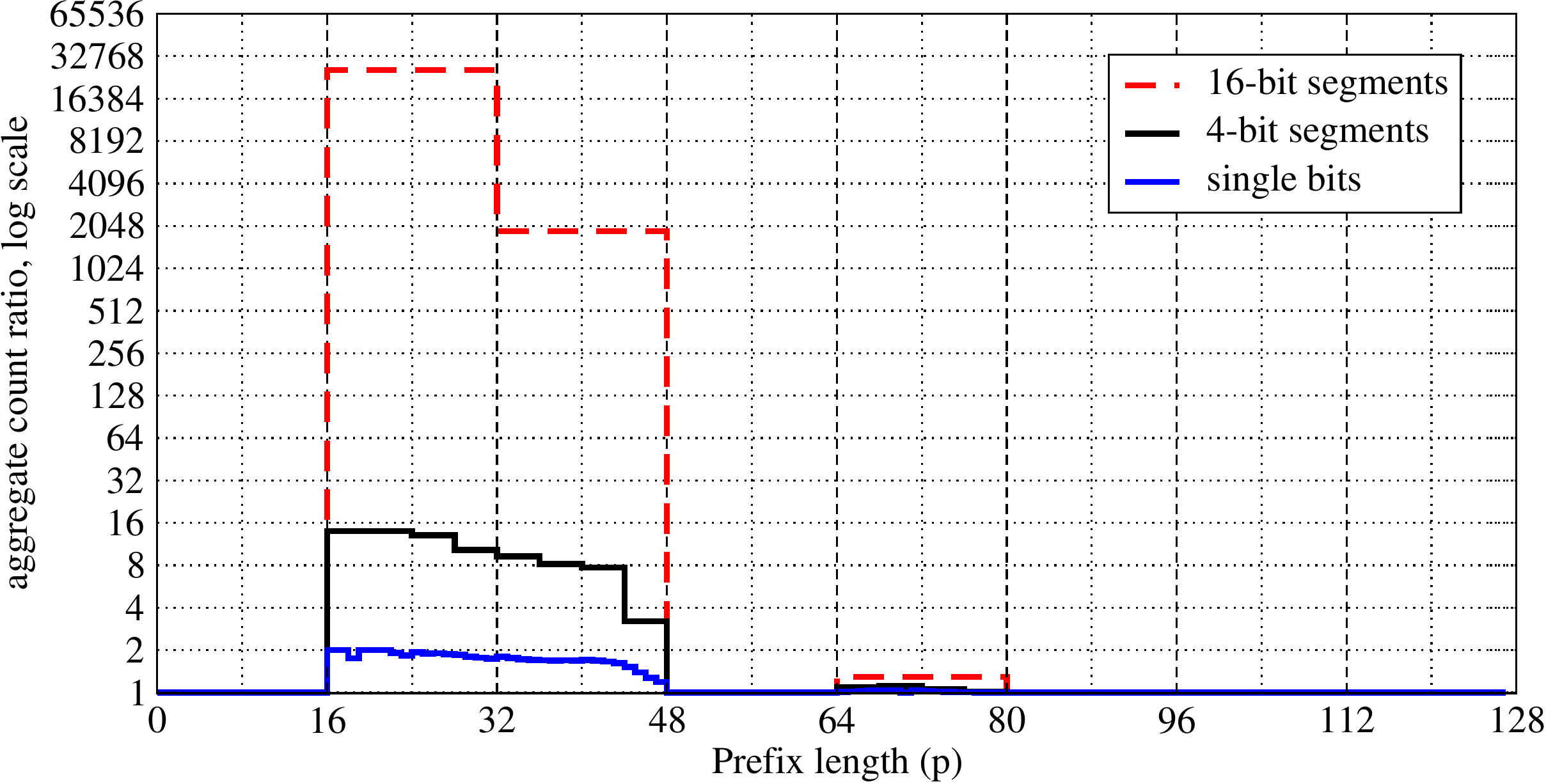}
}
\\
\subfloat[US mobile: 510M active IPv6 client addrs, 167M /64s]
{
\label{fig:USmobileMRAPlot}
\includegraphics[width=0.50\textwidth]{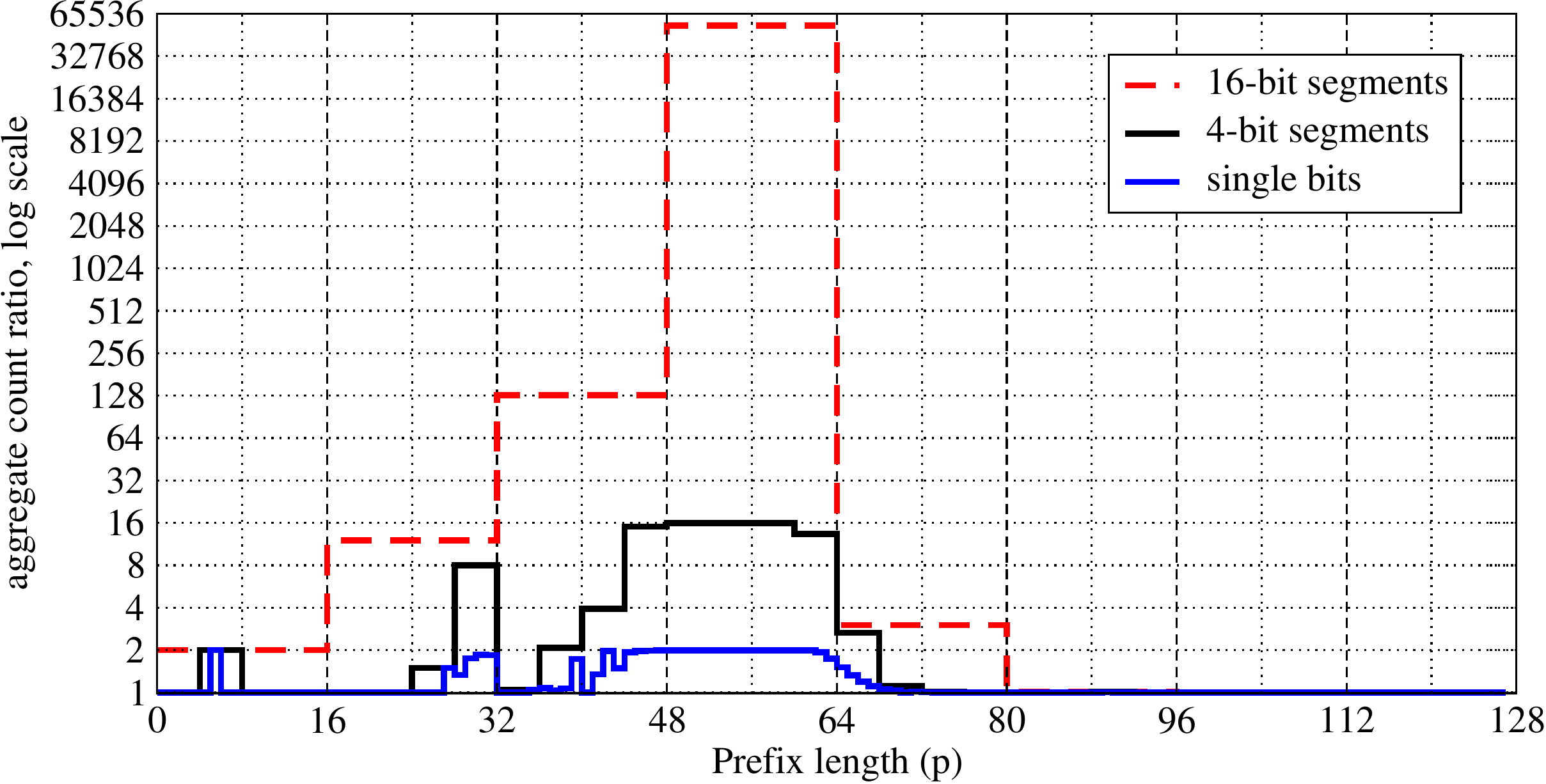}
}
\subfloat[EU ISP prefix: 86.2M active IPv6 client addrs, 15.5M /64s]
{
\label{fig:EUISPMRAPlot}
\includegraphics[width=0.50\textwidth]{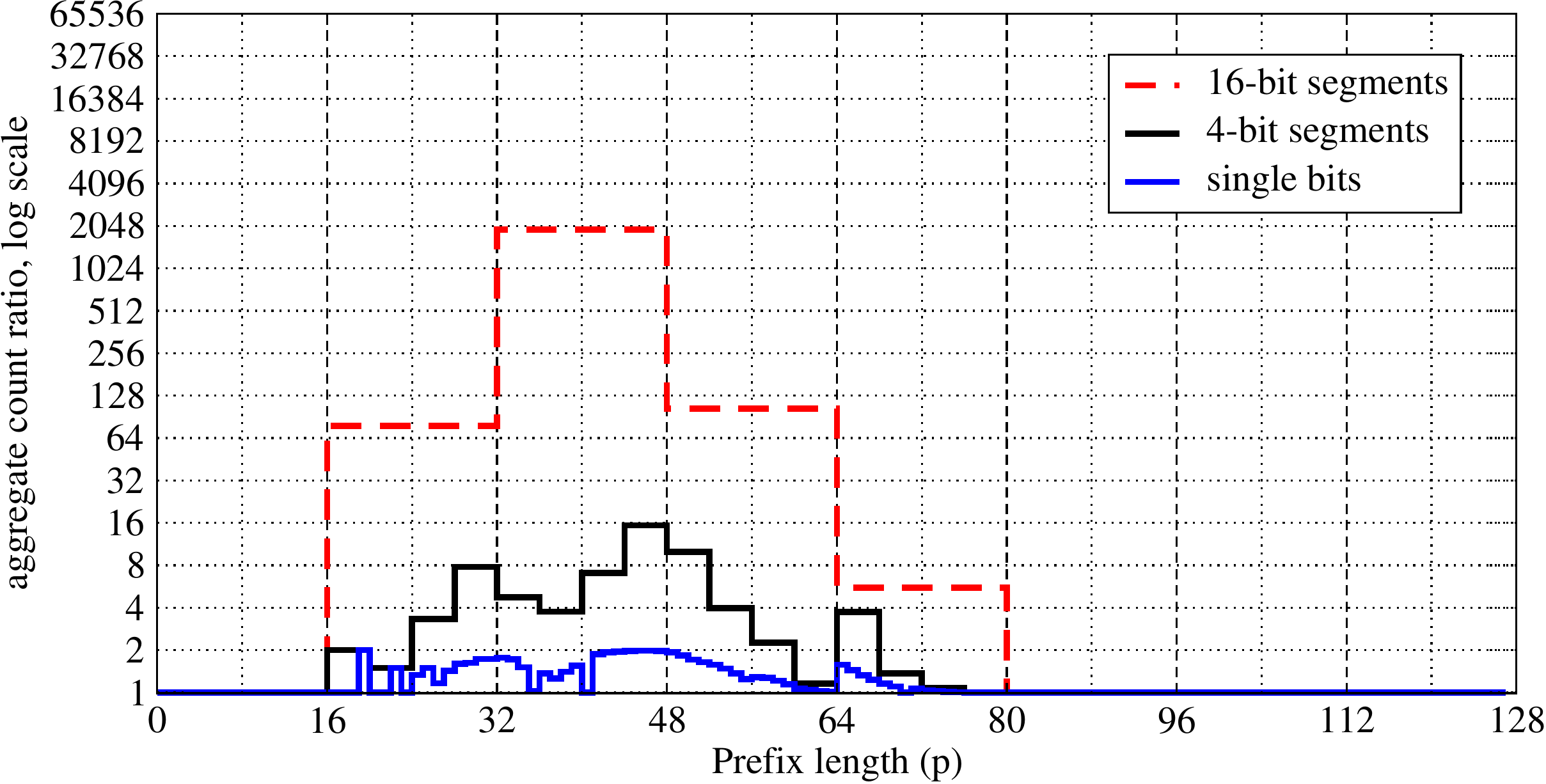}
}
\\
\subfloat[EU univ. dept. prefix: 94 active IPv6 client addrs, 1 /64]
{
\label{fig:EUUniDeptMRAPlot}
\includegraphics[width=0.50\textwidth]{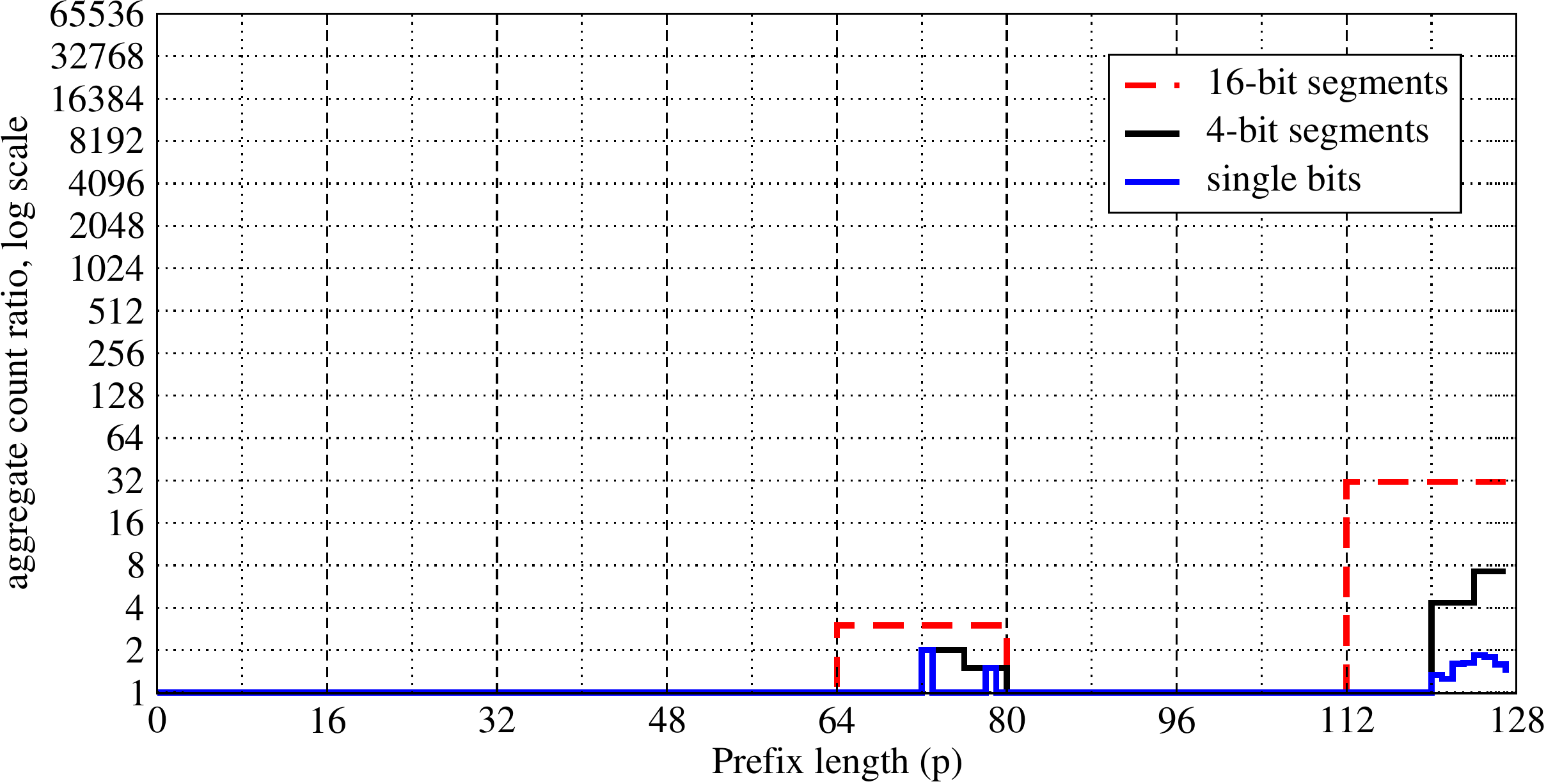}
}
\subfloat[JP ISP prefix: 57.0M active IPv6 client addrs, 2.18M /64s]
{
\label{fig:JPISPMRAPlot}
\includegraphics[width=0.50\textwidth]{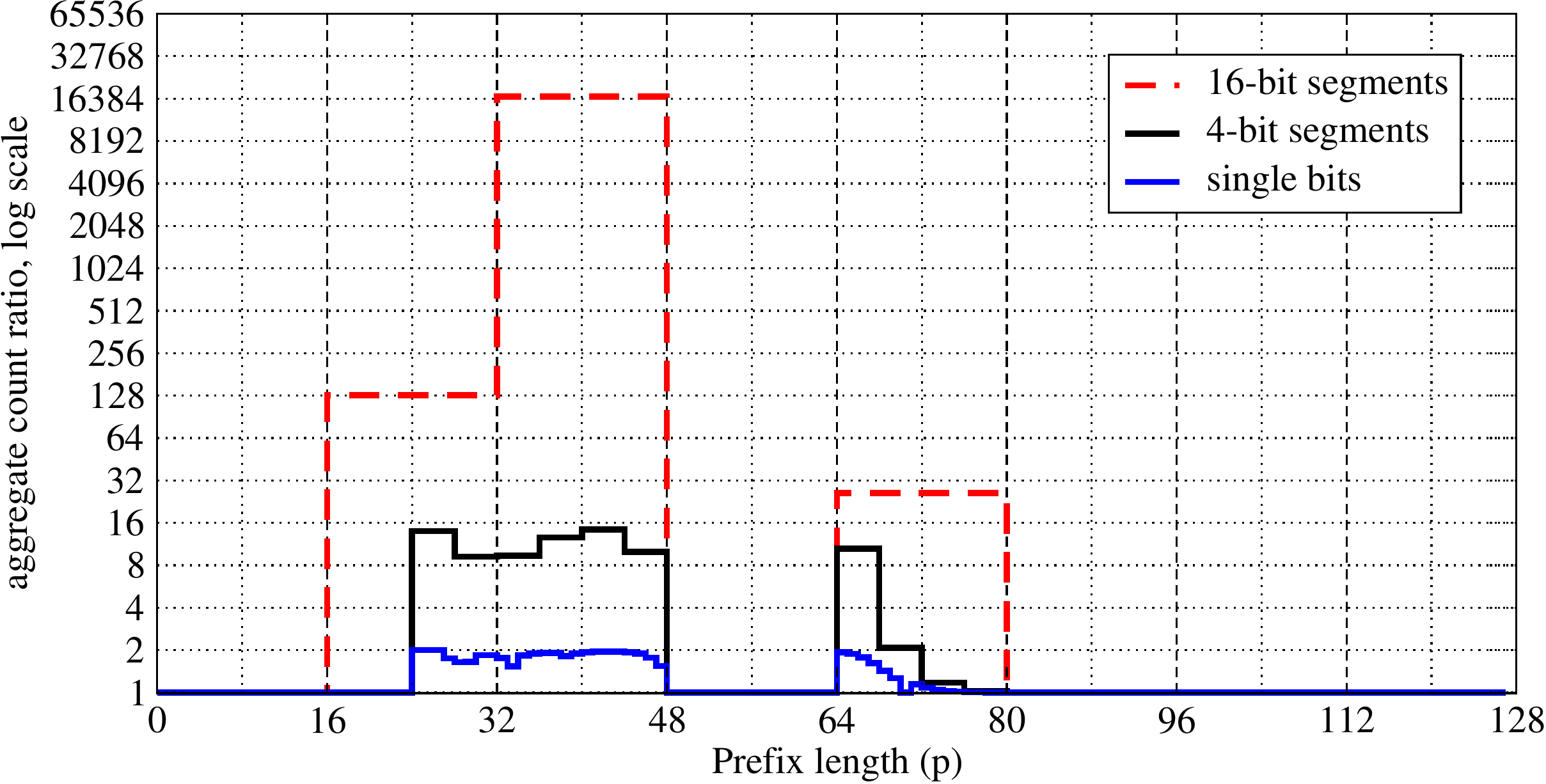}
}
\caption{Distribution and MRA plots for active IPv6 addresses observed during 7 days, March 17-23, 2015.~\label{fig:MRAPlots}}
\end{figure*}

Here we turn to results based on MRA plots that we introduced
in Section~\ref{aggcount}.  With myriad possible MRA plots but
limited space, a data-driven exploration methodology that directs
our attention to ASNs and prefixes of interest is called for.
To this end, we first examine Figures~\ref{fig:bgpCounts}
and~\ref{fig:MRAdist}, distributions of aggregate count ratios
across all active IPv6 ASNs and BGP prefixes.

Consider Figure~\ref{fig:MRAdist}. This is a set of box plots,
each showing the distribution of aggregation ratios across all IPv6 BGP
prefixes for each of the 16-bit segments of each prefixes' set of active
IPv6 addresses.  Unlike a typical box plot showing just the median,
middle 50, and whiskers to, say, the 1st and 99th percentiles, these
also show middle 90\% and whiskers extend to the absolute maximum,
as annotated.  Overall, we can see that most aggregation takes place
across the three 16-bit segments between bits 32 and 80.
We also see that about 20\% of the prefixes (the 75th through
95th percentiles (transparent portion of the box) have significant
aggregation in the 112-128 bit segment, thus we include an MRA plot
for just such a prefix in Figure~\ref{fig:EUUniDeptMRAPlot}.

In Figure~\ref{fig:bgpCounts}, by the solid black line near the lower righthand corner, we see that there is an exceptional ASN
with 500 million active addresses in a week's time, thus we include
its MRA plot as Figure~\ref{fig:USmobileMRAPlot}.  This happens
to be the ASN of the prefix with the highest aggregation in
the 48-64 bit segment in Figure~\ref{fig:MRAdist}.

Figures~\ref{fig:allMRAPlot} through~\ref{fig:JPISPMRAPlot}
are the resulting selected MRA plots for active WWW client
addresses observed March 17-23, 2015.
Let's tour the active IPv6 address space through these plots,
discovering their features of interest.  Coincidentally, the networks
represented in these plots happen
also to be diversely located in the world.  Figure~\ref{fig:allMRAPlot}
is the MRA plot for all active WWW client addresses observed in the
entire IPv6 unicast address space, the proverbial ``30,000
foot view.'' The 0-32 bit segment is roughly governed by the
Reginal Internet Registries (RIRs) via their allocations and assignments to ISPs
and end users (though some allocations are much larger than a /32),
and remaining bits down to /64 are within the area that a network operator uses for subnetting in routing protocols.
Figure~\ref{fig:allMRAPlot} shows that there is greater use of the bit space in the 32-64 range
than the 0-32, with the greatest use of a 16-bit segment in the 32-48 bit range.
Within the 16-32 bit range, the RIRs partition more frequently by the higher-order bits, while in the 32-48 bit range, network operators
partition more frequently by the lower-order bits.
Lastly, the 64-128 bit segment is clearly different, as expected given
the prevalence of ephemeral addresses presumably due to SLAAC and
privacy addressing. While we can't see fine details at this level,
prefix aggregation here happens near bit position 64; this is
because this segment is mostly sparsely populated with random values such that
the majority of hosts' addresses share at most short runs of leading
bits of their IIDs in common with other active addresses in their /64.

Figure~\ref{fig:6to4MRAPlot} is the MRA plot for 6to4 clients. Here
we witness the significant difference between IPv6 and IPv4
aggregation. For addresses in the /16 prefix reserved for 6to4,
IPv4 addresses are embedded in bits 16 through 48, as is clearly
evident in the plot.  (The single bits plotted (blue) in the 16-48 segment are
essentially that which Kohler {\em et al.} studied years ago and
plot in~\cite{DBLP:conf/imc/KohlerLPS02}.)  This 32-bit IPv4 address
segment has much higher aggregation than any similar segments of IPv6
in Figure~\ref{fig:allMRAPlot}.

Figure~\ref{fig:USmobileMRAPlot} is the MRA plot for a U.S.-based
mobile carrier. Its most unusual feature is that the 44-64
bit segment is nearly 100\% utilized when observed over one week's
time. This is evidenced by the 16-bit segments value (dashed red)
and the 4-bits segments value (black) nearly reaching their maximum
possible heights of 64K and 16, respectively. By experiment as a
subscriber, we know
that user equipment (UE) in this mobile service receives a different
/64 prefix on each association, and by comparison to the same plot
over only 1 day (not shown), we can deduce that this network seems
to dynamically assign /64s from pools of addresses in this 44-64
bit segment.  This dynamic assignment has consequences when trying
to estimate subscribers because it can cause the count of active /64s observed
to over-represent the number of subscribers. Corroborating
evidence for a dynamic address component in the 44-64 bit segment is that this
carrier's BGP advertisements consist of over 400 /44 prefixes.
The MRA plot for another top mobile carrier that advertises tens of /40
prefixes (not shown due to limited space) is strikingly similar.

Next, let's consider the MRA plots of a European ISP,
in Figure~\ref{fig:EUISPMRAPlot}, and a Japanese ISP, in
Figure~\ref{fig:JPISPMRAPlot}, for one of each of their advertised
BGP prefixes. The careful observer will note that the prefixes
are at least of size 19 and 24 bits, respectively, as evidenced
by the left-most of the single bits (blue) values. Their numbers
of active addresses are similar and both sets appear to primarily
consist of privacy addresses, sparsely distributed in the 64-128 bit
segment. However, the leading 64-bit portions (left side) of the plots differ
starkly, suggesting very different address plans are in use. Most
notably, in Figure~\ref{fig:EUISPMRAPlot}, the 40-64 bit segment is
populated with many values over a week's time, 
with heavier usage of the higher order bits of this range.
Note that bit 40 seems to be constant and that there is a
subtle perturbation in the single bits (blue) aggregation ratios at
position 56. After examining the distribution (not shown) of values
in bits 40-55, we posit that this segment contains an oft-changing,
pseudorandom 15-bit number beginning at bit 41.  This is followed by
an 8-bit value in bits 56-63 of unknown construction, with all 256
possible values observed, but non-uniform and most often 0x00 or 0x01.
By contrast, in Figure~\ref{fig:JPISPMRAPlot}, the 48-64 bit segment
exhibits seemingly no aggregation, suggesting that each /48 has the
same 16-bit value in every address it contains.
Further, by examining the distribution (not
shown) of /64 counts per IID (or Ethernet MAC address) for the JP ISP's 185K
active EUI-64 addresses, we see that 99.6\% of them were observed in
just one /64 in a week's time; this figure is 67.4\% for the EU ISP.
We discuss a reason for this in Section~\ref{discussSpatial}.

Figure~\ref{fig:EUUniDeptMRAPlot} is the MRA plot for one /64 prefix for
one department at a European university. We selected it for
consideration because it
contains multiple $2$@/$112$-dense prefixes, identified in the forthcoming
results in Section~\ref{densify}.
Consequently, the structure shown is markedly different
from other plots. The WWW client addresses are densely packed,
as evidenced by the values at all resolutions (dashed red, black,
and blue) being most prominent in the 112-128 bit segment; this
indicates that these client addresses are numerically close together,
as one might expect, {\em e.g.} when assigning static addresses
to hosts or when assigning addresses via DHCP.  There don't appear
to be any SLAAC addresses, which require a 64-bit network identifier.
Aggregation seen in the 72-80 bit range and none in bits 80-120
suggests network identifier lengths are between 80 and 120.

\subsubsection{\label{densify}Dense Prefixes and Addresses}

Table~\ref{tab:routerDensityCounts} summarizes the dense prefixes
discovered, by the method described in Section~\ref{computingDense},
using the router addresses dataset described in Section~\ref{sec:data}.
Here we perform a limited search of the parameter space to determine
what combinations of $n$ and $p$ (where $n$ is the number of hosts that must be observed within a
prefix of length $p$ for the prefix to be considered dense) yield
a reasonable number of targets for active measurements.  As we
see, manipulating these two parameters gives significant control over
the prefixes classified as dense and, therefore, the number
of possible target addresses that result.

\begin{table}[ht!]
\scriptsize
\centering
\begin{tabular}{|r||r|r|r|r|} \hline
{Density} & {Dense} & {Router} & {Possible} & {Router Address} \\
{Class} & {Prefixes} & {Addresses} & {Addresses} & {Density} \\
\hline\hline
$2$ @ /$124$ & 43.1K & 116K & 689K & 0.1678459119\\ \hline
$3$ @ /$120$ & 8.28K & 81.0K & {\bf 2.12M} &     0.0382372758\\ \hline
$2$ @ /$120$ & 64.2K & 193K & 16.4M & 0.0117351137\\ \hline
$2$ @ /$116$ & 207K & 568K & 852M & 0.0006670818\\ \hline

$64$ @ /$112$ & 187 & 41.2K & 12.3M & 0.0033593815\\ \hline
$32$ @ /$112$ & 509 & 54.8K & 33.4M & 0.0016417438\\ \hline
$16$ @ /$112$ & 3.06K & 105K & 201M & 0.0005259994\\ \hline
$8$ @ /$112$ & 21.5K & 290K & 1.41B & 0.0002057970\\ \hline
$4$ @ /$112$ & 101K & 681K & 6.63B & 0.0001026403\\ \hline
$2$ @ /$112$ & 367K & 1.29M & 24.1B & 0.0000534072\\ \hline

$2$ @ /$108$ & 289K & 1.72M & 303B & 0.0000056895\\ \hline
$2$ @ /$104$ & 108K & 1.84M & 1.81T & 0.0000010171\\ \hline
\end{tabular}
\caption{Dense prefixes identified at various densities for 3.2M router addrs collected February 2015.
\label{tab:routerDensityCounts}
}
\end{table}

Finally, for active WWW client addresses observed March 17, 2015,
we identify 128 thousand $2$@/$112$-dense prefixes and 1.38 million
WWW client addresses contained therein.
This yields 8.39 billion possible target addresses.
Given that it is feasible to survey the entire IPv4 address
space space by active probing in only minutes~\cite{zmap13}, we propose
that it is similarly feasible to survey these dense regions of the IPv6
address space.  Other IPv6 address datasets could yield
additional sets of dense prefixes to survey.

\subsubsection{\label{discussSpatial}Discussion of Spatial Classification}

To evaluate the interprtation of our MRA plots,
we contacted operators pertaining to networks in Figure~\ref{fig:MRAPlots} and received the following information.
{\em (1)} The high utilization of the 40 - 64 bit address segment
in Figure~\ref{fig:USmobileMRAPlot}
coincides with their subscribers being assigned /64s,
{\em e.g.,} by least recently used,
from a pool sized according to the connection capacity of a gateway.
Thus the /64s are reused by other subscribers. Our results
suggest this reuse can occur in just days.  {\em (2)} The university
of Figure~\ref{fig:MRA_Uni} provided us with their full IPv6 address
plan, and the implication from the figure that we observe only 3 hex
character values matches their address plan.  Two of these indicate ``customer
networks'' and ``large customer networks,'' which are the
portions of their prefix that one would expect to see WWW clients.

After we posited that the IP addresses plotted in Figure~\ref{fig:EUISPMRAPlot} contain a pseudorandom
value in the network identifier, we learned that a European ISP
does just that.  As a supposed privacy-enhancing feature, they allow service
subscribers to have their IP addresses' network identifier changed
on demand, at the press of a button~\cite{DTPrivacyButton2011}.

Regarding the subnet shown in Figure~\ref{fig:EUISPMRAPlot}, we found
a pertinent IPv6 address allocation plan available on the web;
this indicates the university to which the containing /48 is
assigned. Furthermore, we found that {\em every} active address
had an {\tt ip6.arpa} PTR record in the DNS and, thus, were able
to collect names for each of these hosts of which 92 began with
``{\tt dhcpv6-}.''
This is evidence that
the department uses a single /64 to provide IPv6 addresses to a set
of about 100 active hosts. 

We evaluate the application of our dense prefix results by  performing
{\tt ip6.arpa} PTR queries for the 2.12 million possible addresses
prefixes of the $3$@$/120$-dense class, highlighted (bold) in
Table~\ref{tab:routerDensityCounts}. This yielded an additional 47K
domain names more than performing queries for just the active WWW client
addresses. (DNS names are valuable hints to IP geolocation software
because domain names sometimes contain physical location information;
this is especially true for routers.~\cite{Padmanabhan01})

Overall, although our results are based on only a few months of data across
a year-long period, we claim they demonstrate that both temporal
and spatial address classifications can reasonably be performed
at large scale, and that the results are useful in choosing targets for
active measurements and in discovering network-specific addressing practices.

\section{Discussion and Future Work}  			\label{sec:discuss}

\subsection{Counting IPv6}

If one assumes a 1:1 correspondence
between /64 prefixes and IPv6 subscribers or ``user connections,''
the numbers of /64 prefixes are candidate surrogates for IPv6
``user'' counts.  However, this assumption is rather crude.  It is
difficult to say whether numbers of active and stable /64 prefixes
are low or high estimates.  Some networks employ addressing schemes
which cause the count of /64s (active or stable) to overestimate
the number of subscribers, {\em e.g.,} the U.S. mobile carrier
in Figure~\ref{fig:USmobileMRAPlot}. Other networks use
a plan by which the number of active /64s seems a reasonable
estimate of active subscribers, {\em e.g.,} the Japanese ISP in
Figure~\ref{fig:JPISPMRAPlot}. Still other networks place many users
in the same /64, or more specific subnet, causing the count of active
/64s to underestimate the number of user connections, {\em e.g.,}
the network in Figure~\ref{fig:EUUniDeptMRAPlot}.  Evidence shows that the
number active /64s observed in a week's time can miscount IPv6 WWW
client devices by a {\em factor of 100} in either direction.

This challenging situation leads us to conclude that estimating IPv6
user or device counts should be informed by addressing practice on a
per-network or per-prefix basis. This likely requires either inside
information from network operators, or a reliable measurement method
to determine addressing practices from outside.  We've had success
in reverse engineering addressing practice by examining the network
identifiers of EUI-64 addresses over time.  These persistent, unique
IIDs serve as guides that help find our way in areas of the IPv6 address space.

\subsection{Longest Stable Prefixes}

Having achieved some success
reverse engineering network structure ``manually,'' as just described,
we propose that one could automatically discover stable portions of
network identifiers, defined as the set of {\em longest stable prefixes} in a
dataset recording many address observations over time.  By combining
aspects of our temporal and spatial classification techniques, we claim
that it is possible to identify a set of such prefixes, perhaps without
relying on inspection of addresses with long-lived IIDs, {\em e.g.,}
EUI-64. These longest stable prefixes are likely to be significant
aggregates within a network's routing tables, thus this presents
a passive means by which one might glean a network's address plan.
We've begun to explore this prospect and it is a focus of our future work.

\section{Conclusion}                    		\label{sec:sumry}
                                                        In this paper, we present a methodology to classify IPv6 addresses.
We employ two techniques: {\em (1)} temporal analysis to determine
prefix and address stability over time, and {\em (2)} spatial
analysis to determine the structure in which prefixes and addresses
are contained.  We develop classifiers and demonstrate their efficacy
in an empirical study of active IPv6 addresses observed at a large
CDN across a year's time, involving billions of WWW client addresses.
The results of our analyses expose operator addressing practices that
impact the interpretation of Internet measurements. Finally, we
propose that the classifications we develop are applicable,
and likely necessary, to comprehensively survey or census the IPv6
Internet by passive and active means.

\section*{Acknowledgments}				\label{sec:ack}
							We thank Cameron Byrne, Dale Carder, {Pawe\l} Foremski, Jan Galkowski,
Steve Hoey, Geoff Huston, Jeff Kline, Liz Krznarich, David Malone,
George Michaelson, Keung-Chi Ng, and Erik Nygren for their comments and
assistance.

%

%

\bibliographystyle{plain}
\vspace{0.5mm}
\scriptsize
\bibliography{paper}

\end{document}